\documentclass[a4paper,12pt]{article}
\usepackage{amssymb,amsmath,mathbbol,mathrsfs}
\usepackage[usenames,dvipsnames]{color}
\usepackage{hyperref}
\usepackage{xcolor}
\usepackage{stmaryrd}
\usepackage{authblk}
\usepackage{framed}
\usepackage{empheq} 
\usepackage{slashed}
\usepackage{hyphenat}
\usepackage{cite}
\usepackage{apacite}
\usepackage{natbib}
\usepackage{dsfont}
\usepackage{chngcntr}
\usepackage{enumerate}  

\usepackage{marginnote}    

\usepackage[left=.8in,right=.8in,top=.8in,bottom=.8in]{geometry}                  

\usepackage{sectsty} 
\allsectionsfont{\sffamily\mdseries\upshape} 
\usepackage{tocloft}

\makeatletter
\renewenvironment{abstract}{%
    \if@twocolumn
      \section*{\abstractname}%
    \else 
      \begin{center}%
        {\bfseries\sffamily\abstractname\vspace{\z@}}
      \end{center}%
      \quotation
    \fi}
    {\if@twocolumn\else\endquotation\fi}
\makeatother

\numberwithin{equation}{section}
\hypersetup{
	colorlinks=true,         
	linkcolor=MidnightBlue,          
	citecolor=BrickRed,        
	urlcolor=MidnightBlue            
}

\newcommand{\be}{\begin{equation}}
\newcommand{\ee}{\end{equation}}

\newcommand{\F}{{{\Phi}}}
\renewcommand{\d}{{\mathrm{d}}}
\newcommand{\D}{{\mathrm{D}}}

\newcommand{\G}{{\mathcal{G}}}

\newcommand{\pp}{{\partial}}

\newcommand{\Diff}{{\mathrm{Diff}}}

\renewcommand{\bar}{\overline}

\newcommand{\RR}{\mathds{R}} 

\renewcommand{\t}{\mathfrak{t}}

\newtheorem{defi}{Definition}

\newcommand{\cint}{{\int\kern-.87em{<}}}
\newcommand{\sint}{{\int\kern-.75em{\sim}}}
\newcommand{\fint}{{\int\kern-1.00em{\int}}}

\newcommand{\bb}{\mathbb}

\newcommand{\order}[1]{\ensuremath{\mathcal{O}(#1)}}

\let\oldmarginpar\marginpar
\renewcommand\marginpar[1]{\oldmarginpar{\color{red}\raggedright\footnotesize #1}}

\newcommand{\henrique}{\color{RubineRed}}
\newcommand{\old}{\color{red}}

\title{Representational conventions and invariant structure}
\author{Henrique de A. Gomes\footnote{\href{mailto:gomes.ha@gmail.com}{gomes.ha@gmail.com}} \\\it University of Oxford\\ \it Oriel  College, Oxford, OX2 1TQ, UK
}

\begin{document}

\maketitle
\begin{abstract}
  In the philosophical literature, symmetries of physical theories are most often interpreted within the general doctrine called \lq{}Sophistication\rq{}. Roughly speaking, it says that models related by symmetries can peacefully co-exist while representing the same physical possibility. But this interpretation still leaves open two main worries about Sophistication: (a) it allows the individuation of what I call \lq{}structure-tokens\rq{} to remain intractable and thus of limited use, which is why practising physicists frequently invoke  \lq{}relational, symmetry-invariant observables\rq{}; and (b) it leaves us with no formal framework for expressing counterfactual statements about the world. Here I will show that a new Desideratum to be satisfied by theories with symmetries answers these worries. The new Desideratum is that the theory admits what I will call \emph{representational conventions} for its structure-tokens. 

 
 
\end{abstract}
\tableofcontents


\section{Introduction}\label{sec:intro}\label{sec:recap}\label{sec:challenges}


In  \cite{Samediff_1a}, I mounted a defense of Sophistication---an attitude towards symmetry-related models in modern classical field theories---in general terms. When applied to spacetime diffeomorphisms and to gauge symmetries, this defence was construed in terms of  anti-haecceitism about spacetime and  gauge fields. Thus I recovered  the familiar or standard view of physical equivalence between isomorphic models---labeled \emph{Leibniz equivalence} in the philosophical literature (cf. \cite{EarmanNorton1987}).
 Nonetheless, certain questions---or better: worries---about this resolution remained, and here I will answer to some of them.\footnote{I label them \lq{}worries\rq{} and not \lq{}questions\rq{} or \lq{}shortcomings\rq{}. A \lq{}worry\rq{} connotes a temporary challenge: it is a short-lived shortcoming;  and it does not need to be formulated in terms of a question.} 


There are three main `worries' about Sophistication as an attitude towards symmetry-related models that the previous paper \cite{Samediff_1a} left unanswered.  The present  paper aims to  resolve two of these, which are aspects of what I earlier called Worry (2), so I will label them (2a) and (2b):\footnote{I will set aside Worry (3):  as argued forcefully by
 \cite{Belot50}, certain sectors of general relativity, and contrary to Sophistication, there is a general consensus that isomorphisms  relate \emph{different} physical possibilities.  Namely, in the context of spatially asymptotically flat spacetimes, some diffeomorphisms---those that preserve the asymptotic conditions but don't asymptote to the identity map---are interpreted as relating different physical possibilities. The same occurs in Yang-Mills theory \citep{Giulini_asymptotic}. But the reasons given for Sophistication (see \cite{Samediff_1a}) were general; they applied at the level of the whole theory.  So Sophistication, as a doctrine for the whole theory, i.e. all its sectors, seems refuted.}
\begin{enumerate}[(a)]
\item Though Sophistication may give us a handle on  what the underlying ontology of the theory is as a whole, it is unhelpful when we seek to describe individual physical possibilities in an invariant manner. In other words, agreed: Sophistication gives us  a sufficient condition of identification of what two models represent about the world (or contrapositively: a  necessary condition of non-identification). That condition is 
the existence of an isomorphism between the models. (Assuming Criterion (i) of  \cite{Samediff_1a}, the condition is therefore equivalent to the existence of a dynamical symmetry; and  assuming Criterion (ii) of  \cite{Samediff_1a}, such an isomorphism is induced by  an automorphism of the models\rq{} common background structure.)  But this condition falls short in several ways. For  given two models, the sufficient condition is not, \emph{prima facie}, tractable; one cannot try out every diffeomorphism---or, more generally, automorphism of the models' common background structure---to see if one will bring the two models into coincidence. In the mathematical jargon to be developed in Section \ref{sec:rep_gen} (see also \cite[Sec. 2]{Samediff_0}), Sophistication may illuminate the equivalence relation, $\sim$---what I call \emph{the structure-type}---but it does not help to characterise the individual  equivalence classes under this relation, $[\bullet]$---what I call \emph{the structure-tokens}.  Of course, the issue would be solved by a formalism that  describes individual physical possibilities without redundancy, but such a formalism is not required by, and indeed would obviate the need for, Sophistication.  This shortcoming motivates many theoretical physicists to seek relational characterizations of the symmetry-invariant ontology; even while endorsing Sophistication as a general approach (as described in \cite{belot_earman_1999, belot_earman_2001}).  

\item Sophistication's commitment to a strictly qualitative ontology deflates the physical significance  of the set-theoretical elements involved in the usual or standard representation of the models---elements such as the  points representing spacetime events. For instance, once we accept Sophistication\rq{}s sister doctrine of anti-haecceitism and its denial of primitive identity,  we have no notion of correspondence between  points that belong to non-isomorphic spacetimes. But such correspondences are crucial for   counterfactuals to be expressed and interpreted. And beyond counterfactuals, these correspondences are crucial to express superpositions of classical spacetimes, a topic that is currently \emph{en vogue} (cf. \citep{Kabel_et_al2024} and references therein). 

\end{enumerate}

 In this paper it will turn out that answering Worry(2a) in a certain way will also answer Worry (2b). So I begin in  Section \ref{subsec:skeptic_why2} by elaborating (2a). I will there explain why these questions are note merely metaphysical: they are very relevant for current issues in theoretical physics. Then, in Section \ref{sec:rep_conv}, I  introduce the idea of \emph{representational conventions}. Conceptually, such conventions are choices of (idealised) physical relations with which we can completely describe each physical possibility without redundancy; they will suffice to answer both (2a) and (2b). 
In Section \ref{sec:examples} I provide several examples of representational conventions. Focusing on the case in which such conventions are given by choices of gauge-fixing, I will illustrate some of their important properties, such as non-locality. 
 In Section \ref{sec:Healey} I discuss how choices of representational conventions are possible, and how they are made.  
 In Section \ref{sec:rep_app} I will develop the theory of counterparthood based on representational conventions and show that  resolving Worry (2a) via certain kinds of representational conventions also resolves Worry (2b).  That is, the resolution  of Worry (2a) via representational conventions offers not only tractable  conditions of identity for the entire universe, they also provide local correspondence relations between non-isomorphic models.
   And in Section \ref{sec:conclusions} I conclude.

 \section{The challenge of individuating structural content}\label{subsec:skeptic_why2}

Here is how I will organise this Section. In Section \ref{sec:structure_tokens} I will expound Worry 2, which is the focus of this paper, and whose resolution lies beyond what Sophistication can illuminate. In Section \ref{sec:meta} I will relate Worry (2) to issues in the metaphysics of spacetime, but argue that the Worry is not strictly metaphysical: it also casts a shadow on current theoretical physics.  In Section \ref{sec:conditions} I will try to narrow down what could count as a resolution of Worry (2a); this  is where I first introduce \emph{representational conventions}, which will be developed in the rest of the paper.
 
 \subsection{The identification of structure-tokens}\label{sec:structure_tokens}

Even once we have  characterized   to our full satisfaction \emph{the type} of invariant structure instantiated by a theory\rq{}s models we may not have a tractable way to identify the \emph{structure tokens}. That is, we may not have tractable conditions for assessing when two given models are isomorphic. 

This is very familiar in mathematics: the classification problem for a type of mathematical object, sometimes trivially sophomore---e.g. finite-dimensional vector spaces are isomorphic iff they have the same dimensions---and sometimes hard, e.g. seeking all the  topological invariants of topological spaces.\footnote{Note here that \lq{}all\rq{} means that two spaces having the same set of invariants implies that the two spaces are homeomorphic.}


 Agreed, the Sophisticated interpretation---augmented as it was  in \cite{Samediff_1a} by further formal and interpretive desiderata---does provide a sufficient condition for models to  represent the same physical possibility. In a fixed context of application (cf. \cite{Fletcher_hole, Pooley_Read}) this condition amounts to  the existence of an isomorphism between the representing models: this ensures they instantiate the same structure-token.\footnote{Here, I will try to abide by the following terminological distinction: I will call `representation\rq{} the (vexed) relationship between a model and the physical possibility that it is supposed to model;  and I will call `instantiation\rq{} the (less controversial) mathematical relation between the model and the isomorphism-class (the structure-token) that the model belongs to. \label{ftnt:instantiation}} 
 
 An analogy to Leibniz's Principle of the Identity of Indiscernibles (PII) may be helpful here. In order to apply the PII, we cannot check whether Fred and George have \emph{all} the same properties---even all the same qualitative properties! No, instead we apply a tractable condition of identity: whether George and Fred share the same bodies, for instance, or the same exact fingerprints (assuming they can\rq{}t be faked).

 As an example, in general relativity, we may admit that models are Lorentzian manifolds  whose geometric structure is invariant under isometry, and so we have sufficient conditions for representing the same physical possibility---being isometric. But given  two Lorentzian metrics, \emph{when} are they isometric? 
 
  To show that two given models are non-isomorphic, all that Sophistication implies is that if we scan the entire infinite-dimensional space of gauge transformations or diffeomorphisms, and do not find a gauge transformation (respectively, diffeomorphism) between the models, then they are not isomorphic. To be more explicit about this problem:  given two coordinate-based descriptions of metrics  $g^1_{\mu\nu}, g^2_{\mu\nu}$, we would either have to stumble upon a coordinate change that relates them (and so infer that they represent the same physical possibility), or else stumble upon some coordinate invariant function that can be defined on all spacetimes and that takes different values on the two metrics (and so infer that they represent different physical possibilities). And  thus if we fail  on both counts, we could not decisively conclude that the two metrics correspond to same geometry, nor that they correspond to different geometries.

For an illustration of the intricacies of the problem, let us for now focus on the case of Riemannian geometries and take a familiar list of isomorphism-invariant properties. For instance: (1) ``a two-dimensional surface, of area $X$, bounded by two geodesics, of length, $L$''; or  (2) ``the manifold is geodesically convex"; or (3) ``for all points $x$, there exists a unique point $y$ whose geodesic distance $\pi R$ from $x$ is greater than all other points", etc. All of these properties are invariant under isometries. Property (3) is instantiated say, in the two-sphere, where it describes anti-podal points, and it is not instantiated on a plane.  Indeed, it is only property (3) that comes close to fully characterising the geometry: assuming the metric is Euclidean, it would characterise a round sphere of some dimension. And that is only because the round sphere is a highly unusual, highly symmetric model. In contrast, the first two cases  are far from  uniquely  characterising a physical possibility, or even points or regions of the world in which these properties hold. To do so in general we would need an infinite list of properties like (1) and (2). Moreover, it is not clear from inspection when two items in such a list are inconsistent. For instance, in the simple case where (3) is part of the list, the $L$ of item (1) would have to be smaller than $4\pi R^2$. We arrive at this explicit condition because in this simple case (1) fixes the geometry to a large degree, and we can then use familiar theorems of spherical geometry. But in general cases the consistency of a long list would pose an intractable problem: I call this the problem of \emph{consistency} of generic lists of symmetry-invariants; or, more generally, the problem of \emph{consistency of piecewise-invariant descriptions}. We will encounter it again in Section \ref{sec:nonlocal}, when we discuss non-locality. 

The second and main problem that clearly inflicts denumerable lists of geometric invariants is \emph{completeness}. A \emph{complete} set of invariants (also called in the literature on gauge theory a complete set of `observables\rq{}, cf. \citep{HenneauxTeitelboim}) is a set that taken together distinguishes two isomorphism classes of models; viz. distinguishes by a member of the set of observables taking a different value on two models in the two classes. 
But clearly, any \emph{finite}, consistent list of geometric invariants is, generically, \emph{incomplete}. That is because a Riemannian geometry has infinite degrees of freedom. Indeed, the space of Lorentz metrics that satisfy the Einstein equations has non-denumerably many degrees of freedom, i.e. two physical degrees of freedom per space point; and the space of Yang-Mills connections also has continuously many degrees of freedom. This means that, generically, no finite or even denumerable list will completely fix the structure-token for models of a field theory. Thus, generically, any such description can be subject to the problems of incompleteness and inconsistency.\footnote{
And then there is the related worry, about a non-denumerable list of invariants being vastly overcomplete! In this case, the elements of the list could be invariant but, since vastly overcomplete, still have to satisfy constraints that are in practice intractable. So this problem is one of consistency, as above, rather than one of completeness. \label{ftnt:overcomplete}}

Lastly, an attempt to single out a structure token by a  list of invariants brings up the related, but distinct, Worry (2b), also left obscure by Sophistication: how are we  to tractably and qualitatively specify or individuate points, regions, or properties so as to compare them across different physical possibilities? That is, how can we give points and regions, or frames in the gauge case,  consistent conditions of identity that are tractable and sufficient? Here too, we can even take for granted the standard anti-haecceitist argument that one can only individuate an object as a nexus in a web of relations. But that assumption takes us nowhere closer to an exact description of how this individuation is to proceed: what are these relations, and how do they successfully specify, or equivalently, individuate, points? Can they be satisfied in different isomorphism-classes, allowing us to articulate interesting counterfactuals? All we have is   a vague allusion to a conjunction of relations that may well in practice be inconsistent, incomplete, or too rigid to be used in the formulation of counterfactuals.\footnote{Within a single physical possibility, I will use \lq{}specify\rq{} and \lq{}individuate\rq{} synonomously. But `identity\rq{} needs to be construed  more flexibly when we consider different possibilities, and different ways to `identify\rq{} objects across possibilities, in Section \ref{sec:counterfactual}.  In that context I will prefer `specify\rq{} over `individuate\rq{}.\label{ftnt:specify} }   

 
   This is Worry (2b), and this paper will make precise what such relations have to be so that they successfully individuate. And in making it precise, it becomes apparent that there are many different, independent sets of relations that can fully `individuate points as places in a structure\rq{} across possibilities.\footnote{Incidentally, this plurality of sets of relations that are sufficient for individuation is essentially absent from the  allusions to individuation in the literature on symmetry and equivalence. On the contrary, this literature much more often (if not always) considers only the totality of properties and relations in which points stand with respect to each other. But what can we do with that? For instance,  even accepting that  the ontology  is firmly about certain types of relations e.g. distances between spacetime points, and so being clear about the type of structure, we have at first glance no concise way of qualitatively describing how these distances are distributed, since we have no concise  way of qualitatively designating the points that stand in these relations. Here I will propose a more tractable notion of relationism, that further picks out model-dependent subsets of relations.}

\subsection{An  upshot for theoretical physics}\label{sec:meta}
In this Section I will argue that the implications of these Worries are not strictly metaphysical: they are important for current issues in theoretical physics.

 Take the hole argument  (cf. \cite{Pooley_routledge, GomesButterfield_hole1} for recent appraisals).  The argument articulates the threat of a   pernicious form of indeterminism that could arise in general relativity due to the the existence of isomorphic models. 
 Leibniz equivalence and Sophistication stipulate that isomorphic models  represent the same physical situation, and thus eliminate the threat. Of course, the stipulation is not unwarranted: some  judge that it is even  guaranteed by a proper understanding of the mathematical formalism of Lorentzian manifolds (see e.g.   \cite{Weatherall_hole, Fletcher_hole}). 
 
I only partly agree with this judgement. I take the hole argument, at bottom, to point to Worry (2a) about characterising structure-tokens, which is why Einstein\rq{}s `point-coincidences\rq{} arguments is often cited in its resolution.\footnote{The main idea, to be discussed at length throughout the paper,  is to specify points by their web of relations to other points: \lq{}\lq{}All our spacetime  verifications invariably amount to a determination of spacetime coincidences [...] physical  experiences [are] always assessments of point coincidences.\rq{}\rq{} \citep{Einstein_points}.} But, as I said, that general argument is  vague, and does not help to explicitly characterize structure-tokens. Thus, far from being a debate of merely  metaphysical interest,  the  lack of such an explicit characterisation  has been argued to be consequential  for practising physicists, as most extensivelly catalogued by \cite{belot_earman_1999, belot_earman_2001}. They write:
\begin{quote}Far from dismissing the hole argument as a simple-minded
mistake which is irrelevant to understanding general relativity, many
physicists see it as providing crucial insight into the physical content
of general relativity. \cite[p. 169]{belot_earman_1999} \end{quote}

And here is Isham,  on the related difficulty of explicitly characterizing invariant structure (``what constitutes an observable''):
\begin{quote}
The diffeomorphism group moves points around. Invariance under such an
active group of transformations robs the individual points of $M$ of any
fundamental ontological significance . . . [the argument] is closely related to the question of what
constitutes an observable in general relativity---a surprisingly contentious
issue that has generated much debate over the years. \cite[p. 170]{Isham_POT}\end{quote} 
 
 As Isham remarks,  it is a matter of fact that physicists also devote significant attention to finding `observables', which they understand as quantities that are invariant under the symmetries of the theory, that we could use to describe physical content without redundancy.\footnote{See e.g. \citep{Thiemann_2003, Rovelli_book, Donnelly_Giddings} for general arguments surrounding this difficult issue of `gravitational observables'. But there are many examples from the string theory literature as well: e.g. see \citep{Harlow_JT} for an explicit, and rather complicated basis in the simplified context of a two-dimensional gravitational theory called Jackiw-Teitelboim gravity; and \cite{witten2023background} proposes an algebra of operators along an observer's worldline as a background-independent algebra in quantum gravity. \label{ftnt:obs}}  And this matter of fact is hard to reconcile with the idea that Sophistication, or Leibniz equivalence, leaves open no further important questions about symmetry and equivalence once the mathematical notion of isomorphism between models is clarified.

 As to  Worry (2b), resolving it is necessary in order to discuss local spacetime counterfactuals:  modal intuitions---such as that Earth \emph{could have been} closer to the Sun---invoke \emph{some}  relation between spacetime regions across physical possibilities. Since Sophistication denies spacetime points of primitive identity across physical possibilities, we need an alternative way to establish local inter-world comparisons. Indeed, the topic of spacetime counterparts has become even more timely than that of finding invariant observables, since it is germane to current research  on the superpositions of gravitational fields (cf. \citep{Kabel_et_al2024} for a recent appraisal).
 
 `Superspositions of spacetime\rq{} is a quantum theme, which I will mostly steer away from here. But it points to a feature of these worries as they appear for the theoretical physicist: as I will explain in Section \ref{sec:rep_conv}, they only take central stage in theoretical physics in the context of quantization or in the treatment of subsystems.\footnote{Note that, though these arguments and counterarguments have been given within the context of general relativity, there are closely analogous ones for Yang-Mills theory.}   

 \subsection{What are we looking for?}\label{sec:conditions}
 
In Section \ref{sec:challenges}, I reformulated Worry (2) more precisely as stating two shortcomings of Sophistcation. To recap:  Worry (2a) is that we have not  specified how  individual physical possibilities are to be tractably and completely individuated, or uniquely characterised. Worry (2b) is that we have not  provided invariant descriptions of objects within each physical possibility, and thus lack any correspondence---or rather, `counterpart relations'; cf. \cite{GomesButterfield_hole2}---between objects across  possibilities.

Focussing on Worry (2a) in Section \ref{sec:structure_tokens}, I argued that denumerable lists of  geometric invariants  at best incompletely  characterise a structure-token, or at worst characterise nothing at all. So what \emph{would} succeed in answering Worry (2a), by providing consistent representations of structure tokens with tractable conditions of identity? I propose that it is a: 

\begin{defi}[Representational convention]\label{def:rep_conv} A representational convention for a given structure-type is  a complete description for the invariant structure-tokens of the theory such that two such descriptions are identical iff the two structure-tokens are identical. Such a description is a map $\sigma$ from the space of models $\F$ of the theory to a value space $V$, such that, for two models $\phi, \phi\rq{}\in \F$, 
\be\label{eq:sigma_0} \sigma(\phi)=\sigma(\phi\rq{}), \qquad \text{iff}\qquad \phi\sim\phi\rq{},
\ee 
where the equivalence relation is given by isomorphism. 
\end{defi}

The paradigmatically successful example of a characterization that would meet  Worry (2a) and satisfy Definition \ref{def:rep_conv} is, as I briefly mentioned in Section \ref{sec:structure_tokens}, that of  vector spaces, where  specifying the field of scalars and a single number---the dimension of the space---fully characterizes each structure token. But it is not the finiteness of the characterization that makes it successful:  a scalar field  such as temperature or density in a fixed spacetime background could be considered  to be an equally transparent characterization of the ontology. 

In this case of the temperature field suppose there are no isomorphisms apart from the identity. The descriptions are complete and  \emph{local}, in the sense that  two temperature distributions are physically identical iff they numerically match at each point of spacetime. So it is also easy to tractably assess the consistency of partial descriptions.
 But what are good examples of representational conventions for field theories with non-trivial isomorphisms, the cases that this paper focuses on?

 In order to hone  Definitions and find examples,   let us briefly recap both the merits and shortcomings of a description of the structure-tokens via fields that admit non-trivial isomorphisms, for instance, the metric tensors $g_{ab}$. 
One merit is that they provide clearly consistent and complete descriptions of  the invariant structure---e.g. the geometry---of a region where they are defined. 
 But of course,  a metric is not isomorphism-invariant, so it doesn\rq{}t directly provide  tractable conditions for the identity of structure-tokens (as I argued in Section \ref{sec:structure_tokens}, given two metric tensors, it is hard to tell whether they  instantiate the same structure-token, even if it is obvious that each \emph{completely} instantiates \emph{some} structure-token).

And as to Worry (2b),  a  list of invariant geometric quantities suffices to specify  spacetime regions only in very special cases. But from   a generic metric tensor or connection, one could in principle derive   sufficient individuating relations for regions or frames of an associated vector bundle. For instance, generically we could specify each spacetime point by its web of metric relations to other points. But, as with applying the PII for Fred and George in Section \ref{sec:structure_tokens},   it is not clear which tractable subset of relations we should take for this purpose. Moreover, in order to be able to articulate counterfactuals about the contents of spacetime regions in different possibilities, such subsets of relations should not fully determine the content of those regions: we don\rq{}t want spacetime regions to bear their content \emph{essentially}; that would leave no room for interesting counterfactual statements. 

So we are looking for a consistent description of the invariant structure that has the nice properties of the metric and the connection but that can also tractably individuate possibilities and specify regions and frames \emph{across possibilities}.

In field theories,  representational conventions that satisfy these desiderata are most conveniently provided through choices of gauge fixing.   In that case, we require  the output of a representational convention to be itself a model of the original theory, So we define: 
\begin{defi}[Representational convention via gauge-fixing]\label{def:rep_conv_n}
A representational convention via gauge-fixing is a 1-1 map between the equivalence classes given by  isomorphisms of the theory and a subset $\mathcal{F}_\sigma$ of models of the theory, $\sigma: \F\rightarrow \mathcal{F}_\sigma\subset \F$,  such that the two models in the subset are identical iff their arguments are isomorphic; i.e. they obey \eqref{eq:sigma_0}. \end{defi}
Although it can be applied more widerly, this type of representational convention has many advantages particularly in the case of field theories. 

The first is that, for descriptions given through representational conventions satisfying Definition \ref{def:rep_conv_n} the problem of \emph{consistency of piecewise-invariant} descriptions,  as described in Section \ref{sec:structure_tokens}, is tractable. This advantage of Definition \ref{def:rep_conv_n} over  Definition \ref{def:rep_conv} will be cashed out in  Section \ref{sec:consistency}. The second advantage is tightly related to the first: it is that by having structure-tokens instantiated by models of the theory, which are local fields, we can invariantly specify local regions or frames across different physical possibilities, and thus builds a bridge to counterfactuals. It is only for representational conventions satisfying Definition \ref{def:rep_conv_n}  that I can straightforwardly resolve Worry (2b). 

Lastly, I should mention that it is not only gauge-fixings that bear a relationship to my notion of representational conventions, of Definition \ref{def:rep_conv}. In the current literature, they also encompass `relational\rq{} (or `material\rq{}, or even `quantum\rq{}) reference frames, as well as the notion of \emph{dressing}.
And while a construal via gauge-fixing or dressed quantities is most apt to resolve Worry (2a), a construal via relational reference frames is most apt to resolve Worry (2b). We will visit the relationship between these three construals in what follows (see especially Section \ref{subsec:gf}).

In sum:  I will show Worry (2a) is resolved when  we have provided a representational convention for our models that satisfies Definition \ref{def:rep_conv}. As to Worry (2b), I will show it it is partially resolved for representational conventions via gauge-fixing, i.e. satisfying Definition \ref{def:rep_conv_n}.

  \section{Representational conventions}\label{sec:rep_conv}

 Following up on Section \ref{sec:meta}, in Section \ref{sec:rep_phys} I will briefly describe two contexts in which representational conventions take center stage in theoretical physics;  in which, that is, matters of symmetry and invariance are more delicate than most modern literature in philosophy of physics admits: quantization and the treatment of subsystems.  
In Section \ref{sec:rep_gen},  I will introduce the mathematics of representational conventions via gauge-fixing.    In Section \ref{subsec:gf} I will make the relationship  between representational conventions via gauge-fixings, dressings, relational reference frames, and structure-tokens, explicit.  In Section \ref{sec:nonlocal} I will discuss the nonlocality of structure-tokens in the language of dressings and relational reference frames. 

\subsection{Representational conventions in physics: a brief history}\label{sec:rep_phys}

 Section \ref{sec:socio} will explain why representaional conventions are important in quantization, and introduce the idea that gauge-fixings, dressings, and relational reference frames may be different sides of the same coin. Section \ref{sec:subsystems} describes the second context in which representational conventions are important: the treatment of subsystems. Here we will see how a representational convention construed in terms of a reference frame allows us to invariantly describe the physics of subsystems in gauge theory.  

\subsubsection{Gauge-fixing and relational observables in quantization}\label{sec:socio}
In Section \ref{sec:meta}, I argued that theoretical physicists are also preocuppied with matters of symmetry and equivalence; more specifically, with finding complete sets of \lq{}observables\rq{} that fully describe the theory and individuate structure-tokens. But they mostly talk about this in the  context of quantization.  The technical reasons for this   can be very easily illustrated in the path integral, or sum-over-histories approach. There,  the tools of perturbative field theory fail if we do not treat the symmetry-related histories as being, in some sense, physically identical.\footnote{In slightly more detail: since the Hessian of the action vanishes, the propagator around any classical solution diverges, and therefore one obtains divergences at every order of the perturbation series. Finally, these divergences cannot be neatly separated into energy scales and so  they are immune to the machinery of effective field theories.}

There are in general two preferred ways to approach these matters. One approach is to reconstrue the theory, e.g. the path integral, in terms of symmetry-invariant, relational variables: this is the `purist's' approach; common among experts in general relativity. 
In contrast, particle physicists working on gauge theory will not spend much time discussing relational invariant observables. There the usual approach to redundancy is to just ``fix the gauge''. To fix the gauge is to satisfy Definition \ref{def:rep_conv_n} by implementing auxiliary conditions on the representation of the structure-token that are satisfied by a single element in each class of symmetry-related models.  Gauge fixing is  very useful in practice and widely employed in the actual calculations of gauge theory and (classical and perturbative) gravity. But it is also often (erronously) assumed to do violence to the original symmetry of the theory. 

The contrast is clear: in the less applied, more conceptually-driven domains, gauge-fixing  is often overlooked in lieu of a choice of `relational observables' and, more recently, in lieu of  more specific `dressed fields', which is a particular type of relational observable.\footnote{ 
  \cite{Lavelle:1994rh, Lavelle:1995ty, bagan2000charges} were, to my knowledge, the first to explore the relationship between dressings and gauge-fixings, in the context of QCD.}

But I maintain that this assumption,  this contrast sketched above, between gauge-fixing,  relational observables (and reference frames) is in one way misleading. For as we will see in Section \ref{sec:dressing}, gauge fixings can  be understood in terms of symmetry-invariant composites of the original fields, called \emph{dressed fields}, and therefore do no violence to the original symmetry of the theory.  They give rise to a complete set of observables that can be understood relationally, as providing a description of the physical state, or structure-token, with respect to a relational reference frame. (On the other hand, it is possible that a complete set of `observables\rq{} admits no construal in terms of gauge-fixing.)  In Section \ref{subsec:gf},  I will give a conceptually  unified account of dressed quantities, material reference frames, gauge-fixings, and  relational observables.


\subsubsection{A brief history of relational reference frames and subsystems}\label{sec:subsystems}

Apart from the more familiar context of quantisation, discussed in the previous Section, representational conventions in the guise of dressed quantities, sometimes understood as descriptions of a physical state (or structure-token), relative to a relational reference frame, have recently been invoked in the treatment of subsystems in  gauge theory and general relativity. Indeed, our treatment of representational conventions here is inspired by an application to subsystems, which came chronologically first,  in \citep{GomesRiello2016} (where they were described as `abstract material reference frames', or `relational connection-forms'). This initial general treatment was illustrated with explicit examples in \citep{GomesRiello2018, GomesHopfRiello, GomesRiello_new}, where a concept of `abstract reference frames\rq{} was compared to dressings in gauge theory and gravity;  in \citep{GomesStudies, Gomes_new} I gave a  conceptual treatment of these ideas (I\rq{}ll add some detail in Section \ref{sec:obj}). 
  
  
  The reason these notions of dressing or relational reference frames were first introduced in the context of subsystems was that 
  there are subtleties in construing  the isomorphisms  of  subsystem states as dynamical symmetries, subtleties  which can be treated by appealing to  properties of representational conventions.  In particular, there are subtleties about the symmetry-invariance of  a bounded subsystem's dynamical structures, such as its intrinsic Hamiltonian, symplectic structure, and variational principles in general.\footnote{ The point being that dynamical structures on bounded subsystems are \emph{not} invariant under arbitrary gauge transformations of the boundary state, so, in order to preserve gauge-invariance at the level of the subsystems, the boundary state must be understood as `already gauge-fixed\rq{}, even without an explicit gauge-fixing condition. This would in practice fix gauge transformations to the identity at the boundary of a subsystem. In effect, fixing the gauge transformations in this way, for finite boundaries, requires the stipulation of a single value for the fields at the boundary. Of course, this may be very well for a particular subsystem, but it is not a condition we would wish to enforce for \emph{any} subsystem. The procedure is not `subsystem-recursive\rq{}, in the language of \cite{Wallace2019b}. In contrast, a representational convention does not impose any physical limitation on the boundary states, and so is explicitly subsystem-recursive. 
  
  The modern guise of the obstruction to gauge-invariance posed by boundaries was first highlighted in \citep{DonnellyFreidel} (which also introduced `edge-modes' as new degrees of freedom that fixed this obstruction), which caused a flurry of papers on subsystems in gauge theory  (cf. e.g. \citep{Teh_abandon, Geiller:2017xad, DonnellyFreidel, GomesRiello_new, GomesHopfRiello, Riello_symp, GomesStudies, Geiller_edge2020, carrozza2021} and references therein).  See \citep{GomesRiello2016}  for the introduction of relational reference frames as a tool to restore gauge-invariance in a subsystem-recursive manner, and  \citep{carrozza2021} for a more recent and comprehensive review of the relationship between dressings, relational reference frames, and gauge-fixings. \label{ftnt:subsystem}}

  But by describing a model relative to some subset of degrees of freedom, or from the point of view of \lq{}relational reference frames\rq{}, our description is rendered explicitly gauge invariant for all gauge transformations,  even in the presence of arbitrary boundaries.

\subsection{Representational conventions via gauge-fixing: general properties}\label{sec:rep_gen}

At the abstract level explored in this and the next Sections, there are no differences between diffeomorphisms and the symmetries of gauge theories, nor to those of non-relativistic particle mechanics, which mostly lie outside of the scope of this paper but can still be encompassed by the formalism. And so we treat all of these symmetries uniformly, labelling them with the group $\G$, which could be infinite-dimensional. 

Given the space  of models of a theory, ${\F}$, there is an action  of $\G$ on ${\F}$, a map ${\F}:\G\times {\F}\rightarrow {\F}$, that preserves the global structure on $\F$ (e.g. is smooth, in the topology of $\F$), and preserves dynamics, e.g. the Hamiltonian, the action functional, or the equations of motion (in the language of \cite{Samediff_1a}, they are $S$-symmetries).
 More formally: there is  a structure-preserving map, $\mu$, on $\F$ that can be characterized  element-wise,  for $g\in \G$ and $\varphi\in \F$, as follows:
 \begin{eqnarray}\mu:\G\times \F&\rightarrow&\F\nonumber\\
(g,\varphi)&\mapsto&\mu(g, \varphi)=: \varphi^g. \label{eq:group_action}
\end{eqnarray}

The symmetry group partitions the  space of models into equivalence classes in accordance with an equivalence relation, $\sim$, where $\varphi\sim \varphi'$ iff for some $g$, $\varphi'=\varphi^g$. We denote the equivalence classes under this relation by square brackets $[\varphi]$ and  the orbit of $\varphi$ under $\mathcal{G}$ by $\mathcal{O}_\varphi:=\{\varphi^g, g\in \mathcal{G}\}$. Though there is a one-to-one correspondence between $[\varphi]$ and $\mathcal{O}_\varphi$, the latter is rather seen as an embedded manifold of $\F$, whereas the former exists abstractly, outside of $\F$ (which is why I called it `the view from nowhere\rq{}; see Figure 1).
\begin{figure}[h]
  \centering
  \includegraphics[width= 0.5\textwidth]{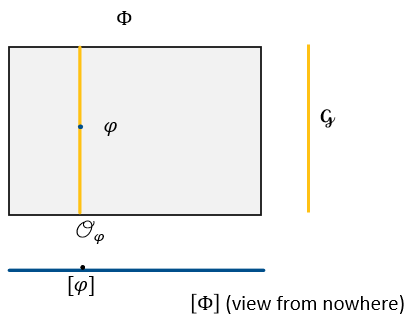}
  \caption{
The space of states, `foliated' by the action of some group $\mathcal{G}$ that preserves the value of some relevant quantity, $S$, and the space of equivalence classes. In field theory, when $S$ is the value of the action functional, the spaces $\mathcal{G}$, $\F$, and $[\F]$ are infinite-dimensional (Frech\'et) manifolds.  
  }\label{fig:PFB_mine}
\end{figure} 
Writing the canonical projection operator onto the equivalence classes, $\mathsf{pr}:{\F}\rightarrow {\F}/\G$, taking $\varphi\mapsto [\varphi]$, then  the orbit ${\cal O}_\varphi$ is the pre-image of this projection, i.e.  $\mathcal{O}_\varphi:=\mathsf{pr}^{-1}([\varphi])$. 

I can now be more precise about what I called \emph{structure-type} and \emph{structure-token}: the structure-type is the equivalence relation $\sim$ and the structure-tokens are the individual $[\varphi]$. It is clear that only knowing the structure-type does not necessarily give us \emph{tractable} conditions of identity for the structure-tokens. But now we will provide these conditions. 

We can now improve on Definition  \ref{def:rep_conv_n} with the more precise:
\begin{defi}[Representational convention via gauge-fixing]\label{def:rep_conv_maths} A representational convention via gauge-fixing is an injective map 
\begin{align}
\label{eq:rep_conv}
\sigma:[\F]&\rightarrow \F\\
[\varphi]&\mapsto \sigma([\varphi])\nonumber
\end{align}
 that  respects the required mathematical structures of ${\F}$, e.g. smoothness or differentiability and is such that $\mathsf{pr}(\sigma([\varphi]))=[\varphi]$, where $\mathsf{pr}$ is the canonical projection map onto the equivalence classes.\label{def:rep_conv_gf}
\end{defi}

Armed with such a choice of representative for each orbit, a generic model $\varphi$ could be written uniquely as some doublet $\varphi=([\varphi], g)_\sigma:=\sigma([\varphi])^g$, which of course satisfies: $\varphi^{g'}=(\sigma([\varphi]))^{gg'}=([\varphi], gg')_\sigma$.\footnote{Our notation is slightly different than \citet[p. 9]{Wallace2019}'s, who denotes these doublets as $(O, g)$ (in our notation $([\varphi], g)$), and  labels the choice of representative (or gauge-fixing) as $\varphi_O$ (our $\varphi_\sigma$). We prefer the latter notation, since it makes it clear that there is a choice to be made.}    Thus we identify ${\F}\simeq [{\F}]\times \G$ via the diffeomorphism:
\begin{align}\label{eq:doublet} \bar\sigma: [{\F}]\times \G&\rightarrow \F\\
([\varphi], g)&\mapsto \sigma([\varphi])^g\nonumber
\end{align}

Now, as I mentioned, the space $[\F]$ is abstract, or only defined implicitly; this is why I called it \lq\lq{}the view from nowhere\rq\rq{}. Since we cannot usually represent elements $[\varphi]$ of $[\F]$ intrinsically, we in practice replace $\sigma$ by an equivalent projection operator that takes any element of a given orbit to the image of $\sigma$: 
\begin{defi}[Projection operator for $\sigma$]
A map 
\begin{align}
h_\sigma:\F&\rightarrow\F\nonumber\\
\varphi&\mapsto  h_\sigma (\varphi)=\sigma([\varphi])\label{eq:proj_h}
,\end{align}
is called the projection operator for  $\sigma$ of Definition \ref{def:rep_conv_gf}. The end result, $h_\sigma (\varphi)$, is called a \emph{dressed} $\varphi$.  \end{defi}
Since $[\varphi^g]=[\varphi]$, we must have
\be\label{eq:h_equiv} h_\sigma(\varphi^g)=h_\sigma(\varphi).\ee
 Now, $h_\sigma(\varphi)$  is an explicit map on $\F$, i.e. it is a function of $\varphi$. So $h_\sigma$ uniquely and concretely represents structural content. Two given models, $\varphi, \varphi'$,  give the same value for \textit{all} symmetry-invariant quantities iff  $h_\sigma(\varphi)=h_\sigma(\varphi')$. 
 
 Thus, using  representational conventions (and the notation $\cal F$ for gauge-fixing) we define:
  \begin{defi}[Structure-tokens according to $\sigma$]\label{def:donder}
     Each structure token $[\varphi]$ is instantiated uniquely, according to $\sigma$, by the corresponding value of the map $h:\F\rightarrow \F$.
     \end{defi}

 \subsection{ Gauge-fixing, dressing, and relational frames in field theories}\label{subsec:gf}

The three Sections below will respectively focus on gauge-fixing, dressings, and relational frames. The common thread in this Section, that I will argue runs through each of these three topics, is the notion of representational convention.

\subsubsection{Gauge-fixing}
  In equation \eqref{eq:doublet}, describing the product structure of the space of models, I oversimplified: finite-dimensional principal bundles need not be trivial, and, accordingly, the space of models need not be globally isomorphic to the product between the space of physical states and the group of gauge transformations. Indeed, even a local product form for the space of models is only guaranteed to exist  in the finite-dimensional case.

 Nonetheless,  it is in fact true that the space of models $\F$---for both Riemannian metrics and gauge connection-forms---is mathematically very similar to a principal fiber bundle, with $\G$ as its structure group (resp. diffeomorphisms and vertical automorphisms). But there are important differences between the infinite-dimensional and the finite-dimensional case. In the finite-dimensional case, it suffices that the action of a group $G$ on the given manifold $P$ be free (and proper) for that manifold to have a principal ${G}$-bundle structure, usually written as ${G}\hookrightarrow P\rightarrow P/{G}$.  In the infinite-dimensional case, these properties of the group action are not enough to guarantee the necessary fibered, or local product structure: one has to construct that structure by first defining a \emph{section}. I summarise the construction of that structure and its main obstructions in Appendix \ref{app:PFB}. 
 
A choice of section is essentially a choice of embedded submanifold on the model space $\F$ that intersects each orbit exactly once. In the 2-dimensional figure 1 above, this would be represented by a `nowhere-vertical' curve: i.e. a submanifold that is transversal to all the $\G$-orbits and intersects each orbit once. 

 That is, we impose further functional equations that the model in the aimed-for representation must satisfy; this is like defining a submanifold through the regular value theorem: in finite dimensions,  defining a co-dimension $k$ surface $\Sigma\subset N$ for some $n> k$-dimensional manifold $N$,  as $\mathcal{F}_\sigma^{-1}(c)$, for $c\in \RR^k$, and $\mathcal{F}_\sigma$ a smooth and regular function, i.e.  $\mathcal{F}_\sigma:N\rightarrow \RR^k$ such that $\mathsf{ker}(T \mathcal{F}_\sigma)= 0$.

Once the surface is defined, $\sigma$ can be seen as the embedding map with range  $\sigma([{\F}])=\mathcal{F}_\sigma^{-1}(0)\subset {\F}$.  The next step is to find a gauge-invariant projection map, $h_\sigma$, that   projects  any configuration  to this surface. 

\subsubsection{Dressing}\label{sec:dressing}

I have called $h_\sigma(\varphi)$ the dressed variables. What are its clothes?  In the case of dressings associated to gauge-fixings, they are the gauge transformation $g_\sigma(\varphi)$  required to transform $\varphi$ to a configuration $\varphi^{g_\sigma(\varphi)}$ which belongs to the gauge-fixing section $\sigma$.\footnote{This is not the most general type of dressing function. Firstly, there is the infinitesimal version, given by the relational connection-form $\varpi$ (cf. \cite{GomesRiello2016, GomesRiello2018, GomesRiello_new} and \cite{GomesHopfRiello}; and \cite{Gomes_new, GomesStudies} and \cite[Appendix A]{GomesButterfield_hole2} for philosophical introductions). Second, there is a more complete theory of dressings and dressed fields, which only requires the right covariance properties and can restrict to subgroups of the gauge group: see  \cite{Francois2021, Francois2021a} for a review. } 

Here we encounter the dual interpretations of gauge-fixings and dressed variables. In the case of fields, the projection $g_\sigma$ is a function of $\varphi$ that is usually non-local and relational, comparing the values of different fields at different points.  Conceptually, the image of this projection---the dressed field---is a    symmetry-invariant composite, or function, of the original degrees of freedom which uniquely describe each structure-token relationally. 
The local interpretation  as a gauge-fixed field has the field  satisfying an auxiliary set of differential equations besides their equations of motion, by belonging to $\mathcal{F}_\sigma$. We can thus see such conditions as implicitly defining the invariant relational composites that we call `dressed variables\rq{}. Let us now see how this definition works. 
 
A function $\mathcal{F}_\sigma$, valued on some general vector space $W$, defining the section (surface) $\sigma([\F])$ as its level surfaces $\mathcal{F}_\sigma^{-1}(0)$ is  suitable as a gauge-fixing  iff it gives rise to a unique dressing by satisfying two conditions: \\
 \noindent$\bullet$\,\,\textit{Universality} (or existence):~ 
For all $\varphi\in \F$, the equation $\mathcal{F}_\sigma(\varphi^g)=0$ must be solvable by a dressing  $g_\sigma:\F\rightarrow \G$.

  That is: 
\be\label{eq:gauge-fixing} g_\sigma:\F\rightarrow \G ,\,\,\text{is such that}  \,\, \mathcal{F}_\sigma(\varphi^{g_\sigma(\varphi)})=0, \,\, \text{for all}\,\, \varphi\in \F.
\ee
This condition ensures that $\mathcal{F}_\sigma$ doesn't impose further physical constraints, i.e. that each orbit possesses at least one intersection with the gauge-fixing section.

\smallskip

\noindent$\bullet$\,\,\textit{Uniqueness}:~ $g_\sigma$ is the unique functional solution of  $\mathcal{F}_\sigma(\varphi^{g_\sigma(\varphi)})=0$. 

From these two conditions, it follows that     $\varphi^{g_\sigma(\varphi)}=\varphi'{}^{g_\sigma(\varphi')}$ if and only if $\varphi\sim \varphi'$, meaning that models have the same projection onto the section iff they are isomorphic. 
In one direction, the claim is immediate: $\varphi^{g_\sigma(\varphi)}=\varphi'{}^{g_\sigma(\varphi')}$ implies $\varphi=\varphi'{}^{g_\sigma(\varphi')g_\sigma(\varphi)}$ which means they are isomorphic. In the other direction, assume $\phi\rq{}=\phi^g$. From \eqref{eq:gauge-fixing}: 
\be \varphi^{g_\sigma(\varphi)}={\sigma}([\varphi]).\ee
Thus, under the group action, $\varphi\mapsto \varphi^g$, and since ${\sigma}([\varphi])={\sigma}([\varphi^g])$ (since $[\varphi^g]=[\varphi]$), we get:
 \be\label{eq:cov_proj}\varphi^{g_\sigma(\varphi)}=(\varphi{}^g)^{g_\sigma(\varphi^g)},\ee which establishes the other direction. And we obtain the following equivariance property for $g_\sigma$:\footnote{These equations are schematically identical to the ones we find for Yang-Mills theory in the finite-dimensional case.  In the Yang-Mills case, a section $\sigma$ in the space of models is essentially a structure-token-dependent section $\sigma$ of the finite-dimensional principal bundle, $P$. }
\be\label{eq:cov_g} g_\sigma(\varphi^g)=g^{-1}g_\sigma(\varphi).
\ee

In possession of a dressing function  $g_\sigma$, it is convenient to rewrite the dressed field $h_\sigma$ of \eqref{eq:proj_h} as $h_\sigma(\bullet):=\bullet^{g_\sigma(\bullet)}$, i.e:
\begin{align}
h_\sigma:\F&\rightarrow\F\nonumber\\
\varphi&\mapsto  h_\sigma (\varphi):=\varphi^{g_\sigma(\varphi)}
\label{eq:h_proj_F}\end{align}
 


And of course, we can still change the representational convention  itself, i.e. act with the group  on the image of $h_\sigma$:  Because $h:\F\rightarrow\F$ is a projection (as opposed to a reduction $\mathsf{pr}:\F\rightarrow [\F]$), and thus $h_\sigma(\varphi)\in \F$, we could consider $(h_\sigma(\varphi))^g$. But the new, transformed model, would no longer obey the conditions defining the previous representational convention: if these conditions are taken to define, for extra-theoretical reasons, `acceptable\rq{} representations,  the new model would not be an `acceptable\rq{} representation of that orbit. 

Given any two representational conventions $\sigma, \sigma'$, we have
\be  \varphi^{g_\sigma(\varphi)}=(\varphi^{g_\sigma\rq{}(\varphi)})^{g_\sigma\rq{}(\varphi)^{-1}g_\sigma(\varphi)}
\ee
and so we obtain
\be\label{eq:transition_h} h_\sigma(\varphi)=(h_{\sigma'}(\varphi))^{\mathfrak{t}_{\sigma\sigma'}(\varphi)},
\ee
where $\mathfrak{t}_{\sigma\sigma'}(\varphi):=g_\sigma\rq{}(\varphi)^{-1}g_\sigma(\varphi)$  is a model-dependent isomorphism (the analog of the transition map between sections of a principal bundle). But, from the covariance of $g_\sigma$, given in \eqref{eq:cov_g}, the transition map is itself gauge invariant, as expected, depending only on the structure-tokens, or orbits $\mathcal{O}_\varphi$.

\subsubsection{Relational reference frames}\label{sec:ref_frames}

Finally, as mentioned in Section \ref{sec:subsystems}, dressed variables have also been related to `material reference frames\rq{}, where `material\rq{} has nothing to do with `matter\rq{}; it is  meant as `physical\rq{}, or `relational\rq{} (I\rq{}ll stick to relational.)  The idea here has a noble tradition, and is a staple of research in quantum gravity (cf. e.g. \citep{Rovelli_book}): it is that we can anchor a particular representation to a physical system, with the ensuing representation being straightforwardly  understood in terms of relations to this physical system. 

For instance, given a set of four scalar fields obeying functionally independent Klein-Gordon equations, we can understand DeDonder gauge as using these fields as coordinates on region of spacetime (see Section \ref{sec:dedonder_gauge}). Similarly, in gauge theory, a particular field could select a particular internal frame for a vector bundle, and we would describe other fields relative to this frame. In electromagnetism, unitary gauge  can be understood in this way (see Section \ref{sec:unitary}). In each of these cases,  $g_\sigma$ can be understood as the transformation required to go from an arbitrary representation of the structure-token to a preferred representation relative to this `relational reference frame\rq{}.  In the case of spacetime, by locating points relative to this physical system, we make precise a frequent claim found in the literature on Sophistication about spacetime, that `spacetime points can only be specified by their web of relations to other points.\rq{}

But this web of relations cannot be too rigid, otherwise each region will bear their specific content essentially.  If the web fixes a region but also entirely fixes the content of that region, no interesting counterfactuals  can be expressed for it. In the context of the hole argument, this kind of rigidity leads to \citep{Maudlin_essence, Maudlin_substance}\rq{}s `metric essentialist\rq{} resolution, which has been criticised for being an argument valid only for the `actual world\rq{} \citep{Butterfield_hole}; and in \citep{GomesButterfield_hole2}  precisely for not allowing the expression of counterfactuals. 

  The use of relational reference frames to solve Worry (2b), about counterfactuals, is thus contingent on their association with a gauge-fixing. We must use just enough relations between parts of the fields to fix the representation but not limit the physical content, and this is guaranteed by  the \emph{Uniqueness} and \emph{Existence} properties of a gauge-fixing.\footnote{Indeed, it is for this reason that relational reference frames associated to gauge-fixings are able to restore a gauge-invariant notion of subsystem-recursivity: see footnote \ref{ftnt:subsystem}.}

However, if we  associate `relational reference frames\rq{} to gauge-fixing in this way, we impose a constraint on what kind of  reference frames we are countenancing: only those that are related by the symmetries of the theory, as illustrated by  \eqref{eq:transition_h}. This is not a bad thing, since we should expect that systems described by reference frames not thus associated would not obey the same laws. 

We will see specific examples of reference frames associated to gauge-fixings in Section \ref{sec:examples} (in particular, in Sections \ref{sec:unitary} and \ref{sec:dedonder_gauge}), and we will employ this construal explicitly in resolving Worry (2b) in terms of counterpart theory in Section \ref{sec:counterfactual}.

In sum, for fields,  the existence of  a gauge-fixing section means $\varphi$ and $\varphi'$   instantiate the same structure and so represent the same physical possibility iff the dressed field values match at \emph{every} spacetime point. Thus, for each choice, a dressed field over a certain region gives a complete physical description of that region, without redundancy, and which can be interpreted relationally. 
The projection \eqref{eq:proj_h} reduces  problems of  identity of physical states to numerical identity over spacetime.

\subsection{Non locality}\label{sec:nonlocal}

This Section deals with aspects of non-locality associated to dressings and gauge-fixings. In Section \ref{sec:dressing_ctraints} I will discuss the relationship between dressings, constraints, and non-locality. The kind of non-locality that emerges via dressings in gauge theory forces us to again face   the question of \emph{consistency} of partial descriptions of a structure-token. This question was  introduced  in Section \ref{sec:structure_tokens}, where I argued that descriptions via a denumerable list of invariants was generally inconsistent.  In Section \ref{sec:consistency}, I will show that, for representational conventions via gauge-fixing, the question of consistency of piecewise-invariant descriptions, described in Section \ref{sec:structure_tokens}, has a neat answer. 

\subsubsection{Dressing and constraints }\label{sec:dressing_ctraints}
In  Section \ref{sec:examples}, I will give examples of dressings obtained via representational conventions, most of which are gauge-fixings.  And most, whether obtained via a gauge-fixing or not, will display some degree of non-locality. This feature is expected for generic states of  field theories with symmetries. 

For instance, even without appealing to a gauge-fixing, it is well-known that diffeomorphism-invariant functions are usually non-local: this is a notorious problem for quantum gravity (see \cite{Thiemann_2003, Rovelli_book, Donnelly_Giddings}). Thus \citet[p. 170]{Isham_POT} continues the quote of Section \ref{sec:meta}:
\begin{quote} In the present
context, the natural objects [that constitute `observables' in general relativity] are $\Diff(M)$-invariant spacetime integrals [...]
Thus the `observables' of quantum gravity are intrinsically non-local.\end{quote}

Why is it that dressed quantities are generally non-local? Let me try to give a brief conceptual explanation.  A local symmetry implies, via Noether\rq{}s second theorem (cf. \cite{BradingBrown_Noether} for a conceptual exposition), that the equations of motion of the theory are not all independent, and thus they only uniquely determine the evolution of a subset of the original degrees of freedom. There is thus a certain freedom in choosing which `components of the fields' will be uniquely propagated, or will evolve deterministically---each choice corresponds to  \emph{a choice of gauge}.

For general relativity and Yang-Mills theory, such choices of initial data must satisfy  elliptic equations, whose solutions require integrals over an initial spacelike surface.
  As I  argue at further length in \cite{Samediff_2}, in practice, ellipticity means that boundary value problems require only the boundary configuration  of the field, i.e. they  do not also require the field's rate of change at the boundary. Thus the solution of these equations exists \emph{on each simultaneity surface}; and so the solution  does not describe the propagation of a field, as would a solution to a hyperbolic equation. Thus non-locality arises because the function that takes the original local degrees of freedom to this uniquely propagated subset of local degrees of freedom is, generally, non-local: the value of an element in the subset at point $x$ depends on the values of the original degrees of freedom at other points.  
This (classical and non-signalling) non-locality can be quantified to a certain extent, cf. \cite{Gomes_new, GomesButterfield_electro} for a more comprehensive discussion of this point.\footnote{ But  it is clear that relativistic causality holds for the isomorphism-invariant facts, since (quasi-)hyperbolicity of the equations of motion ensures causality is respected for one choice of metric or gauge potential representing the isomorphism-equivalence class  (e.g. Lorenz or DeDonder gauge for electromagnetism and general relativity, respectively).}

 In sum,  gauge-fixings or dressed representations generally require this particular (rather benign) type of holism, or non-separability, that applies to theories with elliptic initial value constraints, such as general relativity and Yang-Mills theories; but does not apply to theories like Klein-Gordon in a fixed Minkowski background, nor to any `gauge theory' whose gauge transformations involve no derivatives (see Section \ref{sec:unitary} and \cite[Sec. 1.1, point (3)]{GomesButterfield_electro} for more discussion about this sort of (non-signalling)   classical non-locality.) 

I\rq{}ll finish this Section with a quote from \cite[p. 3]{Harlow_JT}, who,  in a recent  paper pursuing a  symmetry invariant description of a particular  gravitational  theory, nicely summarise the relationship between dressings, constraints, gauge-invariance, and relationism:
\begin{quote}
It has long been understood that diffeomorphism symmetry must be a gauge symmetry, and that physical observables must therefore be invariant under almost all diffeomorphisms. [...] There is an analogous situation in electromagnetism: fields which carry electric charge are unphysical unless they are “dressed” with Wilson lines attaching them either to other fields with the opposite charge or to the boundary of spacetime. One way to think about dressed observables in electromagnetism is that they create both a charged particle and its associated Coulomb field, which ensures that the resulting configuration obeys the Gauss constraint. Similarly in gravity any local observable by itself will not be diffeomorphism-invariant, so we must dress it [...creating] for it a gravitational field that obeys
the constraint equations of gravity. In practice such observables are usually constructed
by a ``relational” approach: rather than saying we study an observable at some fixed
coordinate location, we instead define its location relative to some other features of the
state.
\end{quote}

 \subsubsection{Consistency of partial descriptions and gluing: the advantage of gauge-fixing}\label{sec:consistency}
 
  But a question remains: how does non-locality, such an ubiquituous feature of dressings, differ from the type of holism illustrated by denumerable lists of geometric invariants, that will often lead to an inconsistent description, as discussed in Section \ref{sec:structure_tokens}?
I will show that, in the case of dressings obtained via gauge-fixing of field theories, the consistency of piecewise-invariant descriptions  is straightforward. 

Briefly, suppose two regions, $R_+$ and $ R_-$  (or $R_\pm$ for short) overlap on a region $R_0:=R_+\cap R_-$. Further suppose that the union of all three regions is a manifold, and that each region is an embedded submanifold therein. Consider the fields intrinsic to each region, and their  intrinsic isomorphisms; call these $\F_{\pm}, \F_0$ and $\mathcal{G}_\pm, \mathcal{G}_0$, respectively.    For given gauge-fixings $\sigma_\pm$ of $\mathcal{G}_\pm$ on $R_\pm$, we have $h^+(\varphi^+)\in \F_+, h^-(\varphi^-)\in \F_-$,   which we again call $h^\pm$ for short. Then $h^\pm$ are consistent iff given a representational convention $\sigma_0$ (not necessarily via gauge-fixing) on $R_0$, 
\be\label{eq:gf_overlap} h^0(h^+_{|R_0})=h^0(h^-_{|R_0}),
\ee
where $h^\pm_{|R_0}$ represents the restriction of  the $h^\pm$ to the region $R_0$, and so $h^\pm_{|R_0}\in \F_0$.  Equation \eqref{eq:gf_overlap}  means that the $h^\pm_{|R_0}$ are related by a (structure-token-dependent) transformation $\frak{t}_\pm$, as described in \eqref{eq:transition_h}. Thus $h^\pm$ are consistent, or can be `glued\rq{}, or composed into a single global model, iff \eqref{eq:gf_overlap} holds. 

  For a local dressing, $h(\varphi)(x)=h(\varphi(x))(x)$. So, for the same representational convention chosen throughout the manifold: 
\be h^+(\varphi_{|R_+})=h(\varphi)_{|R_+}, \quad \text{thus} \quad h^+_{|R_0}=h_{|R_0}= h^-_{|R_0}.\ee
So, although \eqref{eq:gf_overlap} applies to both local and non-local dressings, it is only non-trivial in the non-local case. 
In the non-local case, even if the choice of gauge-fixing on $R_+$ was the same as that in $R_-$ (e.g. Coulomb gauge),  $h^\pm$,  when restricted to $R_0$, may differ.\footnote{The non-locality of dressings is equivalent to the non-commutativity between regional restrictions and the  dressing map. That is, suppose we are given a representational convention $\sigma$, which is applicable to any region, e.g. by stipulating a gauge-fixing and boundary condition. Suppose further that the dressing function requires solving an elliptic equation (as in e.g. Coulomb gauge, cf. Section \ref{sec:Coul}). Then for a generic submanifold $R$, with boundary $\partial R$:
\be h(\varphi)_{|R}\neq h(\varphi_{|R}).
\ee
This is easy to see in terms of boundary conditions: although $h(\varphi_{|R})$ will necessarily obey the stipulated boundary condition on $\partial R$, the values of $h(\varphi)_{|\partial R}$ will be rather arbitrary, and highly dependent on which surface we choose. And the difference won\rq{}t be constrained to $\partial R$ of course, since solutions of elliptic equations are globally dependent on the boundary condition. This is discussed at lengh, including complications about whether the region-intrinsic isomorphisms are also dynamical symmetries of the theory,  in \citep[Ch. 6]{Gomes_PhD}. }
 And yet, even if they differ over that region, their restrictions may still instantiate the same structure-tokens for that region. This is why, in the non-local case, criterion \eqref{eq:gf_overlap}---that the dressed quantities $h^\pm$ were also models of the theory---was important: in order to assess consistency of the piecewise dressed quantities, these quantities must enter as arguments of the representational convention $h_0$. When $\sigma_0$ is also a representational convention via gauge-fixing, the consistency condition given in  \eqref{eq:gf_overlap} amounts to solving two differential equations and checking for numerical coincidence.  This is an advantage of   representational conventions via gauge-fixing, as per Definition \ref{def:rep_conv_gf}, over other kinds of representational conventions. 
 
 Indeed,  along the lines of an argument originally made by \cite{RovelliGauge2013} in the case of particles and different sets of fields, in \citep{GomesStudies} it was argued  that the importance of gauge is tied to non-locality, since `gluing\rq{} or composing gauge-invariant quantities in the non-local case requires us to match different representations of the same structure-tokens, and to do that, we need transformations  that function like gauge transformations: the $\mathfrak{t}_{\sigma\sigma'}$, as given in \eqref{eq:transition_h} (see also footnote \ref{ftnt:Rovelli}).  The `gluing\rq{} and composition of gauge-fixed quantities is described in detail for Yang-Mills theory in \citep[Sec. 6]{GomesRiello_new}.

\section{Examples  of Representational Conventions}\label{sec:examples}

Representational conventions can be very general. Here I will provide several examples, mostly based on gauge-fixings, starting from non-relativistic particle mechanics to gauge theory and general relativity. 
\subsection{Non-relativistic particle mechanics}

\subsubsection{Center of mass for non-relativistic  point-particles}\label{sec:NRPM}
Take Newtonian mechanics for a system with  $N$  particles, written in configuration space as trajectories $\varphi(t)=q_\alpha(t)\in \RR^{3N}$ (with $\alpha=1, \ldots, N $ labeling the particles), and with shift symmetry:
\be\label{eq:particle_sym}
\varphi^g = (q^\alpha (t) + g), \quad \text{with } g \in \mathbb{R}^3.
\ee
Now we take the section $\mathcal{F}_\sigma$ to be defined by the center of mass:
\be\label{eq:com} \mathcal{F}_\sigma(q) = \sum_\alpha q^\alpha m^\alpha = 0.\ee

The most straightforward way to find the projection $h(q)$, is displayed in \eqref{eq:h_proj_F}. It requires us to find the dressing as a functional of the configuration. That is, take
\be h^\alpha (q) = q^\alpha + g_\sigma(q),\ee
and solve
\be  \mathcal{F}_\sigma(h(q)) = 0 \quad \text{ for } g_\sigma: \Phi \to \mathbb{R}^3 \,\, \text{and arbitrary } \,\, q.\ee
We obtain:
\begin{eqnarray*}
\sum_\alpha (q^\alpha +g_\sigma(q)) m^\alpha= 0 
\Rightarrow g_\sigma(q) = -\frac{\sum_\alpha q^\alpha m^\alpha}{\sum_\alpha m^\alpha}.
\end{eqnarray*}
Clearly, 
\be g_\sigma(\varphi^g) = g_\sigma(q + g) = g_\sigma(q) - g.\ee
Thus if follows that the new set of configuration variables, written as functionals of arbitrary configuration variables, is gauge-invariant (now including indices for both particles and components in $\RR^3$, to be completely explicit): 
\be h_i^\alpha (q^\alpha + g) = (q_i^\alpha (t)) + g + g_\sigma(q) - g = h_i^\alpha (q^\alpha).\ee
And it is also easy to see that, if $h^\alpha (q) =h^\alpha (q\rq{})$, then $q^\alpha-q\rq{}^\alpha$ is given by a shift, and so  $h^\alpha (q) =h^\alpha (q\rq{})$ iff $q$ and $q\rq{}$ are isomorphic. 

This example is still non-local, in the sense of Section \ref{sec:nonlocal}: given two particle subsystems, indexed by $N_+, N_-\subset N$, with $N_+\cup N_-=N$,\footnote{ I am here denoting a set of particle labels with $N_\pm, N$, whereas before $N$ was a number. But this slight innacuracy does not justify the amount of notation that would need to be introduced in order to clarify this point.} the center of mass of either subsystem is generically not identical to the center of mass of their union. But, in the notation of Section \ref{sec:consistency}, for $N_0=N_+\cap N_-$, it is straightforward to verify that, although 
$h^+_{|N_0}\neq h^-_{|N_0}$ (since the particles in $N_0$ are being described relative to  different centers of mass), if $N_+, N_-\subset N$, with $N_+\cup N_-=N$, equation \eqref{eq:gf_overlap} will hold, since both sides correspond to the description of the $N_0$ subsystem with respect to its own center of mass.\footnote{Of couse, we need not have $N_+, N_-\subset N$, with $N_+\cup N_-=N$. That is, we need not have $h_+, h_-$ coming from the restrictions of a consistent global model: equation \eqref{eq:gf_overlap} is supposed to assess consistency, not assume it. If  
$h_0(h^+_{|N_0})\neq h_0(h^-_{|N_0})$
 there is  a bijection between the labels of the particles in $N_0$ according to $N_-$ and $N_+$, but the intrinsic statess of the restricted subsystems are not symmetry-equivalent.}  


Of course, examples abound: beyond the case of translations and boosts,  we could fix a frame in $\RR^3$  by diagonalizing the moment of inertia tensor around the center of mass; in this case $g_\sigma$ would be a state-dependent rotation (see \cite[Sec. 4]{DES_gf} for details).\footnote{Configurations that are collinear, or are spherically symmetric, etc. would be unable to fix the representation: these are \emph{reducible} configurations; they have stabilizers that cannot be fixed by any feature of the structure-token. \label{ftnt:collin}} 

\subsubsection{Inter-particle distances}\label{sec:inter}

This is the first example of a representational convention which is \emph{not} equivalent, at least \emph{prima facie}, to any gauge-fixing. We take the same system as in the previous example, described in \eqref{eq:particle_sym}. And instead of \eqref{eq:particle_sym}, we take, following \cite{RovelliGauge2013}, a new complete set of relational variables: 
\be \bar q_\alpha(t):=(q_{\alpha+1}-q_\alpha), \quad \text{for} \,\, \alpha=1, \cdots, N-1.
\ee
Thus the new variables provide a shift-invariant description of the system. It is clear that they are complete, since the dimension of the space of configurations of $N$ particles in e.g. $\RR^3$ is $\RR^{3N}$, and modulo translations, this is $\RR^{3(N-1)}$. So they satisfy Definition \ref{def:rep_conv}. But there is a clear disadvantage to this kind of description:  because the invariant variables are not elements of the original space of models, $\F$, it is not a representational convention in the sense of Definition \ref{def:rep_conv_n} (or the more precise Definition \ref{def:rep_conv_maths}). So we cannot iteratively apply representational conventions in order to assess consistency, as in \eqref{eq:gf_overlap}.\footnote{Suppose we were coupling two subsystems for which $N_+\cap N_-=\emptyset$.  Now we have $\RR^{3(N_\pm-1)}$ degrees of freedom in each subsystem. But, after coupling, we have $\RR^{3(N_++N_--1)}\neq \RR^{3(N_+-1)}+\RR^{3(N_--1)}$.  Indeed, the coupled system has 3 more degrees of freedom than the individual subsystems. \citet{RovelliGauge2013} argues that these are the gauge degrees of freedom that need to be retained for the coupling of subsystems. \citet{GomesRiello_new}  show that the same can be applied to gauge theories. The difference is that if we restrict the division of the fields to supervene on a complete division of the manifold by complementary regions, the number of extra, holistic degrees of freedom is the dimension of the stabilliser group of the states at the boundaries between the complementary regions.  See \cite{Gomes_new} for a conceptual exposition. \label{ftnt:Rovelli}}

\subsection{Gauge theory}
\subsubsection{Electromagnetism: Coulomb and Lorenz gauge}\label{sec:Coul}

  The Lorenz gauge in electromagnetism is explicitly Lorentz covariant (and has its virtues thoroughly extolled by \cite{Mattingly_gauge}, who argues  that it should be considered as \cite{Maudlin_response}'s ``ONE TRUE GAUGE'').  \cite{Maudlin_ontology} himself endorses a different choice, Coulomb gauge.
  
  But the two gauges are similar: the main difference is that the Lorenz gauge concerns all the components of the potential in a Lorentzian manifold, whereas Coulomb gauge concerns only spatial components of the gauge potential, so is described in a Riemannian manifold. But here I will ignore the difference by construing Lorenz gauge in a spacetime manifold with  Euclidean signature, and Coulomb gauge to be a choice of gauge in the Hamiltonian version of electromagnetism.\footnote{The difference matters: the Lorenz choice in a Lorentzian manifold gives rise to a hyperbolic, and Coulomb always gives rise  to  an elliptic, partial differential equation. In the Lorentzian case with a Lorenz choice, the Maxwell equations of motion for $A_a$ become the  hyperbolic equation (which gives a well-posed initial value problem: see \cite{Samediff_2} for more details about the relation between the IVP and gauge freedom):
$ \square A_a=0,$
This equation still has the gauge-freedom corresponding to $\phi$ such that $\square g=0$. To fix $g$ uniquely one requires strict boundary conditions, which could still carry the non-local behavior we have alluded to in Section \ref{sec:nonlocal} (cf. \cite{Samediff_2}). Thus, in the Lorentzian setting, this gauge-fixing is not complete and we would require additional imput about the initial state. In contrast, the kernel of the elliptic equation $\nabla^2 g=0$ corresponding to Coulomb gauge can be defined in the absence of spatial boundaries:  it is zero. So Coulomb, but not Lorenz, define a \emph{bona-fide} gauge-fixing condition without the need of auxiliary conditions. \label{ftnt:Lorentz}} 

This particular example of $\mathcal{F}_\sigma$ (and   $h_\sigma$, and $g_\sigma$) in field theory is fully worked out by \cite{GomesButterfield_electro}, alongside its physical interpretation. This choice arises from a split of the electric field into a component that is purely Coulombic, or determined by the synchronic distribution of charges, and another component that is purely radiative.  And, apart from worries with \emph{Uniqueness} (described in Section \ref{sec:dressing}, associated to the Gribov problem (discussed in Appendix \ref{app:PFB}), there are straightforward extensions of these gauge-fixings to the non-Abelian case.

In the Coulomb case for Hamiltonian electromagnetism, the field \(\varphi\) is a doublet that transforms only in one component, i.e. 
\be
\varphi = (E^i, A_i), \quad \varphi^g = (E^i, A_i + \nabla_i g),
\ee
with \(g \in C^\infty(M)\), where \(M\) is a spatial (Cauchy) surface, $E_i$ is the electric field, and $A_i$ is the spatial gauge potential, with spatial indices $i, j$ etc. The Lorenz case would be similar, but $M$ would be the spacetime manifold and we would have: 
\be
\varphi =  A_\mu, \quad \varphi^g =  A_\mu + \nabla_\mu g,
\ee
with $\mu, \nu$, etc, spacetime indices. In the following, I will subsume both cases using the abstract index notation $A_a$.

First, following the definition of the dressed function, in \eqref{eq:h_proj_F}, we write: 
\be\label{eq:h_gen_A} h(\mathbf{A})_a:=A_a+\nabla_a g_\sigma(\mathbf{A}),
\ee
where $g_\sigma: \F\rightarrow C^\infty(M)$, and $M$ is either the space or the spacetime manifold (with Euclidean signature), and $\F$ is the space of gauge potentials over $M$ satisfying appropriate boundary conditions (which I will not here discuss: for more details, about boundary conditions and the point that this gauge is not complete in the Lorentzian signature: see footnote \ref{ftnt:Lorentz} and  \citep{GomesRiello_new}).  The condition to be satisfied by $g_\sigma$ will be obtained from our definition of $\mathcal{F}_\sigma:\F\rightarrow C^\infty(M)$, with:
\be\label{eq:div_free}\mathcal{F}_\sigma(\mathbf{A})=\nabla^a A_a=0.\ee
That is, we obtain:
\be  \mathcal{F}_\sigma( h(\mathbf{A}))=\nabla^a(A_a+\nabla_a g_\sigma(\mathbf{A}))=0,\ee
and thus:
\be g_\sigma(\mathbf{A})=-\nabla^{-2}\nabla^b A_b,\ee
 which clearly satisfies the covariance equation \eqref{eq:cov_g}, namely: 
 \be\label{eq:cov_A} g_\sigma(\mathbf{A}^g)=-\nabla^{-2}\nabla^b (A_b+\nabla_b g)=g_\sigma(\mathbf{A})-g.
 \ee 
Since the inverse Laplacian is determined only up to a constant, so is $g_\sigma$, but this is a happy case in which the degeneracy in $g_\sigma$ is a stabiliser of the gauge potential (cf. Appendix \ref{app:PFB}), and so has no effect on the dressed variable. 

The corresponding dressed functional is:
\be\label{eq:h_A} h(\mathbf{A})_a:=A^{g_\sigma(A)}_a=A_a-\nabla_a g_\sigma(A)=A_a-\nabla_a(\nabla^{-2}\nabla^b A_b)\ee
where  $\nabla^{-2}$ is the inverse Laplacian (in the Lorentzian case, given sufficiently strict boundary conditions, it would be a Green\rq{}s function, $\square^{-1}$, with $\square$ the D\rq{}Alembertian). 
Gauge-invariance of the dressed function is implied by \eqref{eq:cov_A} and \eqref{eq:h_A}.
And of course, from \eqref{eq:h_gen_A}, it is clear that if two dressed potentials match, they are dressings of gauge-related potentials. 
  Thus two dressed variables match iff what is being dressed is related by an isomorphism.
 
 Note that, consistently with the duality between gauge-fixing and dressing,  one can describe $h(A)_a$ by describing a non-trivial partial differential equation that it satisfies, or one can provide an unconstrained 1-form $A_a$, but then $h(A)_a$ is seen as a non-local dressed field. Are all such dressed fields non-local in this way?

\subsubsection{Maxwell Klein-Gordon: unitary gauge}\label{sec:unitary}
This gauge is only available for  some sectors of Abelian theories like electromagnetism with a nowhere vanishing charged scalar (Maxwell Klein-Gordon). As we will see, this choice of gauge is \emph{sui generis} for being local (see \cite{Wallace_unitary} for an in-depth analysis of the merits and shortcomings of this gauge).

The Maxwell-Klein-Gordon theory has models \(\varphi = (A_\mu, \psi)\), with $A_\mu$ the gauge potential and $\psi$ a complex scalar function. The gauge transformation of the models is defined as:
\be
\varphi^g = (A_\mu + \partial_\mu g, \psi e^{ig}),
\ee
with \(g \in C^\infty(M)\), and \(M\) representing spacetime.

In the unitary gauge, we have the condition
\be\label{eq:unitary}
\mathcal{F}_\sigma(\varphi) = |\psi| - \psi = 0.
\ee
The dressed variables are:
\be\label{eq:dressed_unitary}
h(A, \psi) = (A_\mu + \partial_\mu g_\sigma(\psi), \psi e^{i g_\sigma(\psi)}),
\ee
where we solve \(F(h(\varphi)) = 0\) for \(g_\sigma: \Phi \to C^\infty(M)\) and arbitrary \(\varphi\).
Assuming \(|\psi|(x) \neq 0, \forall x \in M\), we write:
\be
\psi = |\psi| e^{i\theta}, \text{ with } \theta = -i\ln \frac{\psi}{|\psi|}.\footnote{There is an issue here that the logarithm of complex functions is not well-defined over the entire $\bb C-\{0\}$, because of the periodicity of solutions. Thus we should choose a subset of $\bb C-\{0\}$ that contains a single branch; but nothing depends on the branch.}
\ee

Finally, \eqref{eq:unitary} and \eqref{eq:dressed_unitary}  imply that:
\be
|\psi| e^{i(\theta + g_\sigma(\psi))} = |\psi|,
\ee
so we find
\be
g_\sigma(\psi) = -\theta, 
\ee
which clearly has the right gauge-covariance properties. 
Thus,
\be
h(A, \psi) = h(A, |\psi|) = (A_\mu - \partial_\mu \theta, |\psi|),
\ee
which is  gauge-invariant. But here $A_\mu - \partial_\mu \theta$ is an unconstrained, invariant 1-form $\tilde A_\mu$,  so we rewrite the dressed fields as an unconstrained doublet: $(\tilde A_\mu, \rho)$, where $\rho\in C_+^\infty(M)$ (it is an everywhere positive smooth scalar function) and $\tilde A_\mu\in C^\infty(T^*M)$. So this is an explicitly local representation of the structure-token; in particular, the consistency of piecewise-invariant descriptions amounts to numerical coincidence.\footnote{A natural extension of this unitary gauge to the non-Abelian case takes a nowhere vanishing matter field to define an internal direction in the typical fibers of the matter sector; and, by extension, it defines a gauge-fixing for every other field that couples to this one. In the context of covariant perturbative gauge-fixings, this was labeled `the Higgs connection\rq{} in \citep[Sec. 7]{GomesHopfRiello}.   In the case of SU(2), a closely related decomposition of an arbitrary field using an everywhere non-zero internal direction is known as the The Cho-Duan-Ge (CDG) decomposition \citep{CDG}. It was rediscovered at about the turn of the century by several groups were readdressing the stability of the chromomonopole condensate: the ensuing  decomposition of the gauge potential helps in identifying the topological structures like monopoles and vortices in the gauge field configuration (see \citep{CDG_rev} for a more recent concise summary of the literature and an application to QCD). }

  In closing, this list of gauge-fixings for Yang-Mills theory is of course not exhaustive. There are many other, physically motivated choices we could have made. For instance, we may also want to highlight the helicity degrees of freedom of the theory, in which case we would use temporal gauge.
But temporal gauge, like the Coulomb  one, depend on a foliation of spacetime by spacelike surfaces, and so, strictly speaking, cannot be given a univocal physical interpretation unless  coupled to a representational convention about spacetime foliations. We turn to this in Section \ref{sec:GR}.  Before getting there, I will now exhibit a choice of dressed variables, that, like the choice of inter-particle distances for non-relativistic mechanics described in Section \ref{sec:inter}, is \emph{not} obtained by a gauge-fixing section.

\subsubsection{Holonomies}
A holonomy basis for Abelian gauge theories associates to each loop in spacetime a gauge-invariant phase, namely: 
\be\label{eq:hol} hol_\gamma(A) =\exp{i\int_\gamma A}, \quad\text{for each}\quad \gamma:S^1\rightarrow M.\ee  
More generally, the holonomy thus defined is Lie-group-valued, and, in the non-Abelian case, it is thus not entirely gauge-invariant, but transforms covariantly in the adjoint representation. Nonetheless, there is a straightforward way to regain gauge-invariant variables by taking the trace of the holonomy: these are called  \emph{Wilson-loops}.

The Wilson-loop basis for gauge-invariant quantities is a representational convention in the sense of Definition \ref{def:rep_conv}, but not in the sense of \ref{def:rep_conv_gf}: although they `dress\rq{} the original variables, they do not take values in the original space of models (they take values in the space of real-valued loops). Thus Wilson-loops do not have the two alternative interpretations described for gauge-fixings (as local functions satisfying certain differential equations, or as unconstrained non-local functions of the original variables).  In particular, the  consistency of the composition of piecewise-invariant descriptions (discussed in Section \ref{sec:structure_tokens} in the case of denumerable lists of invariants), cannot be settled by a simple criterion such as \eqref{eq:gf_overlap}, which applies only for representational conventions via gauge-fixing. 

In more detail: unlike dressed quantities obtained from gauge-fixings, the gauge-invariant variables derived from  holonomies is vastly overcomplete. Thus the composition of these variables must obey certain constraints (cf. footnote \ref{ftnt:overcomplete}). And whereas  the consistency of the composition of piecewise-invariant descriptions via dressed variables obtained via gauge-fixing is settled by  numerical coincidence of solutions of partial differential equations (cf.  \eqref{eq:gf_overlap}), the constraints to be satisfied by the composition of holonomies are known as \emph{Mandelstam identities}, and they   can be rather complicated in the case of non-Abelian gauge theories  (I briefly discuss this in Appendix \ref{app:hol}). Thus, in order to assess the consistency of piecewise invariant descriptions via holonomies, i.e. to know whether all the values of the Wilson-loops can be obtained from the holonomies of a single gauge potential, we need to be able to solve these constraints. This is, again, if not intractable, much harder than solving the differential equations in the gauge-fixing case.\footnote{This is not to mention the non-separability of holonomies: even in the Abelian case, the holonomies of region $A$ and of region $B$ do not fully determine the holonomies of region $A\cup B$ if this union has a non-trivial topology. }

Here we see complications for describing gluing or composition for representational conventions according to Definition \ref{def:rep_conv} (but not satisfying Definition \ref{def:rep_conv_gf}).  Relatedly, it is unclear whether a Wilson-loop basis could be used in order to resolve Worry (2b).Going in the other direction, if one assumes that all Wilson loops already satisfy the Mandelstam constraints, it \emph{is} possible to reconstruct the corresponding structure-token, cf. \citep{Barrett_hol}.

\subsection{General relativity}\label{sec:GR}\label{sec:dedonder_gauge}

In the case of Lagrangian general relativity, I will start with De Donder gauge, which fixes coordinates so that the densitized metric is divergence-free: 
\be \mathcal{F}_\sigma(g)=\pp_\mu(g^{\mu\nu}\sqrt{g})=0.\ee
 Rewriting this equation  as $\square x^\mu=0$, one can see De Donder gauge as anchoring  coordinate systems on waves of massless scalars: four different solutions of the relativistic wave equation. I will not bore the reader with yet another demonstration that this gauge gives rise to dressed gravitational fields with the expected properties, described in Section \ref{subsec:gf}.\footnote{As in footnote \ref{ftnt:active_passive}, we are here also in effect assuming the active-passive correspondence of \cite[Sec. 5]{Samediff_0}, to say that fixing a choice of coordinates is equivalent to picking out a unique model within an isomorphism class. 
   In fact, as in the case of Lorenz gauge in gauge theory (cf. footnote \ref{ftnt:Lorentz}), De Donder or harmonic gauge does not completely fix coordinate freedom in general relativity.  Given the coordinates in an initial surface, the gauge uniquely defines coordinates to the past and future of that surface. Moreover,  the use of coordinates implies that in general such conditions are only local.  Giving a global, or geometric version of deDonder gauge can  accomplished using an auxiliary metric, a measure of distance between metrics, and a diffeomorphism that is used to compare them (see \cite[Sec. 7.6]{Landsman_GR}). 
 This strategy---of using auxiliary metrics---is often employed in converting  results originally expressed in particular coordinate systems into explicitly covariant results.
 } 
 
In the corresponding spacetimes, given  coordinates in an initial hypersurface $\Sigma$ (or a portion thereof), we \emph{define} coordinates  in some region of spacetime using four non-local scalar functionals of the metric, that solve four wave-equations:
\be\label{eq:Komar_coords}\square \mathcal{R}^\mu[g_{ab}; p)=0\ee
where I used DeWitt\rq{}s mixed functional notation: the variables depend functionally on the metric (possibly non-locally), but take values locally on the manifold. 
In a less coordinate-centric language, the idea here is  to specify spacetime points through a certain choice of their qualitative properties. That is, we define point $p_x(g_{ab})$ as the point in which a given list of scalars $\mathcal{R}^{\mu}[g;p)$, $\mu=1, \cdots, 4$ takes a specific list of values, $(x_1, \cdots, x_4)$'). In other words, fixing the metric and initial values, the four scalar quantities, $\mathcal{R}_g^{\mu}$  define a map: 
\begin{equation}\label{eq:map_R}
   \mathcal{R}_g= (\mathcal{R}_g^{1}, \cdots \mathcal{R}_g^{4}):M\rightarrow \RR^4.
\end{equation} 
 To pick out points $p\in M$ by the value of the quadruple we invert the map \eqref{eq:map_R}.  
Assuming that the map \eqref{eq:map_R} is  a diffeomorphism---in general it is only locally one---there is a unique value, for all of the models, of $g_{ab}(\mathcal{R}_g^{-1})(x)$, for any $x\in \RR^4$.\footnote{More commonly, for each $g_{ab}$, there will be only a subset $U\subset M$ that is mapped diffeomorphically to $\RR^4$.\label{ftnt:U}} That is, applying the chain rule for the transformations of  $\mathcal{R}_g$, 
\begin{equation}\label{KKS}
 \forall f\in Diff(M), \quad  g_{ab}\circ \mathcal{R}_g^{-1}=f^*g_{ab}\circ \mathcal{R}_{f^*g}^{-1}.
\end{equation}
So, using this quadruple, we have a unique representation of the metric on $\RR^4$. Given some metric tensor $g^{\kappa\gamma}$ in coordinates $x^\kappa$, we can compute the metric in the new, $\mathcal{R}^\mu$ coordinate system as:
\be\label{eq:Komar_trans} h^{\mu\nu}=\frac{\pp  \mathcal{R}^\mu}{\pp x^\kappa}\frac{\pp  \mathcal{R}^\nu}{\pp x^\gamma}g^{\kappa\gamma}.
\ee
This is just a family of 10  scalar functions indexed by $\mu$ and $\nu$: the left-hand-side is the dressed variable, $h_\sigma^{\mu\nu}$, and the $\mathcal{R}^\mu$ are the dressing functions.

Of course, equations \eqref{eq:map_R} and \eqref{KKS} would hold for any of the other choices obtained from a viable gauge-fixing, since those choices also describe a local physical coordinate system, or a relational reference frame, on a patch of spacetime. 
And indeed, any such relation specifying points in terms of their qualitative properties explicitly furnishes a specific counterpart relation  across  isomorphic and non-isomorphic models; cf.  \citep{GomesButterfield_counter}.\footnote{ \citet[p. 468]{Curiel2018} construes qualitative identity of points similarly: \begin{quote}
Once one has the identification of spacetime
points with equivalence classes of values of scalar fields, one can as easily say
that the points are the objects with primitive ontological significance, and the
physical systems are defined by the values of fields at those points, those values
being attributes of their associated points only per accidens.
\end{quote} But he does not construe diffeomorphisms as intra-theoretic changes of convention about the choices of scalar fields, as we do.   }

 And there are many such examples. One that is widely used to study black hole mergers, and initial value problems in general, is  `CMC gauge' (cf. e.g. respectively \citep{Pretorius_2005,York}): this choice adjusts clocks and simultaneity surfaces so that simultaneous observers measure the same local expansion of the universe. In this case, $\mathcal{F}_\sigma(\gamma_{ij}, \pi^{ij})=g^{ij}\pi_{ij}=$const, where $\gamma_{ij}$ is the spatial metric and $\pi^{ij}$ is its conjugate momentum, obtained in the Hamiltonian (3+1) formulation of general relativity \citep{ADM}.

 Another example that does not require any fields other than the gravitational ones, is known as \emph{Komar-Kretschmann} variables. For this example, we must first restrict our attention to spacetimes that are not homogeneous, i.e. generic spacetimes (i.e. excluding Pirani's type II and III spaces of pure
radiation, in addition to excluding symmetric type I
spacetimes). Once this is done, we consider  $ \mathcal{R}^{\mu}(g_{ab}(p)),\,\, \mu=1, \cdots 4$, formed by certain real scalar functions of the Riemman tensor. Since the spacetimes considered here are suitably inhomogeneous, they all contain points in which these functions are linearly independent, and so can be used to specify location without limiting the physical content of the spacetime region  \emph{as a whole}.\footnote{ \cite{Komar_inv} finds these real scalars through an eigenvalue problem: 
 $$(R_{{a}{b}{c}{d}}-\lambda(g_{{a}{c}}g_{{b}{d}}+g_{{a}{d}}g_{{b}{c}}))V^{{c}{d}}=0,$$
 where $V^{{c}{d}}$ is an anti-symmetric tensor. The requirement ensures solutions $\lambda$, whose existence we assume, are independent real scalar functions. \citet[p.1183]{Komar_inv} takes these scalars to be preferred, ``since they are the only nontrivial scalars which are
of least possible order in derivatives of the metric, thus
making them the simplest and most natural choice.'' }
 This choice is exceptional for not requiring auxiliary fields, or spacetime curves or points: the relational reference frame that it provides supervenes only on properties of the metric tensor.\footnote{See \citep{BamontiRF} for a classification of different types of reference frames in general relativity, according to their coupling to the metric and to the inclusion of back-reaction.}

But given particular spacetime points or curves---taken to have, so to speak, a \emph{rigid designation}, across a suitably restricted set of models---we can find adapted coordinate systems or reference frames. Perhaps the most famous choice of coordinates is \lq{}Riemann-normal\rq{} coordinates, obtained by applying the Riemann exponential map for the tangent space of a given event.  This choice is useful if we fix a material system whose scale is small compared to the curvature scale: in these coordinates the metric is described as almost flat along the trajectory of the system; the Riemann curvature appears only at second order in the proper distance to the freely-falling trajectory of the material system.

 Another common example, indeed the most applied on everyday navigation on Earth, is a coordinate choice based on GPS sattelites. This requires four sattelites (taken to be small, and non-back-reacting, and thus following timelike geodesics), which  cross at an initial event and emit timed light-like signals. Those signals cover a spacetime region, and their values provide it with a coordinate system.  (cf. \citep{Rovelli_GPS} for details).  
   Similarly to GPS coordinates, the `dressed' diffeomorphism-invariant observables of \cite{Donnelly_Giddings} are  anchored on non-null spacetime curves.


\section{How to choose a representational convention? }\label{sec:Healey}


In the previous Section, I exhibited several possible choices of representational conventions via gauge-fixing. How should we choose amongst them? Before answering the question of this Section, I will motivate it with  an argument from \cite{Healey_book}.  
  More specifically,  I  take issue with one argument in Healey’s (otherwise outstanding!) book on the philosophical interpretation of gauge theories. It occurs in his Chapter on classical gauge theories (Chapter 4).
Here, by answering Healey, I will   (superficially) connect the ideas  to a popular topic in analytic philosophy: functionalism.

I will start in Section \ref{sec:healey_intro} by recapitulating Healey\rq{}s argument. 
Then in Section  \ref{sec:antiheal} I will show  that, \emph{contra} Healey, intra-theoretic resources enable us to pick out representational conventions for  gauge theory.



\subsection{Healey's argument from functional roles}\label{sec:healey_intro}


\cite[Section 4.2]{Healey_book}\rq{}s  describes  what he takes to be a fundamental difference between the tenability of gauge-fixing in general relativity and in Yang-Mills theory. That is, a difference between specifying, amongst the infinitely many physically equivalent representatives, a particular spacetime distribution of the gauge potentials or of the metric. As I understand Healey, he posits that this specification is easy for the metric, but impossible for the gauge potential. It is easy for the metric because  Lewis\rq{}s ideas about functionalism apply to it, but, allegedly, they don\rq{}t apply to gauge potentials. And with this alleged contrast I will disagree. My disagreement will shed light on how representational conventions are chosen. 

To spell out Healey's  argument in more detail, I will now indulge in a bit of  `Healey exegesis'. 
Healey admits that within a theory there may be many terms standing for unobservable items, such as those variables that are not gauge-invariant. But unobservability by itself is not bad news, since  \citet{Lewis_defT, Lewis_func}'s construction, which I will summarize shortly below,  applies  to the observable-unobservable distinction. It allows us to specify the meaning of unobservable items via their relations to observable ones and to each other. The idea is that we can simultaneously specify what several theoretical  items---labeleled T-terms---refer to,   by their each uniquely satisfying some description (usually called ``functional role'') that can be formulated in terms of each other and of  the better understood---perhaps even observable---terms, labeled O-terms.  

But  how could we use the theory at hand to functionally single out any item if the theory  treats it and others, at least in certain respects, as being on a par, i.e. related by some symmetry? As described by   \citet{LewisRamsey}, the response  is to appeal to patterns of facts   of ``geography'' to break the underdetermination:
\begin{quote}
Should we worry about symmetries, for instance the symmetry
between positive and negative charge? No: even if positive and negative
charge were exactly alike in their nomological roles, it would still be true
that negative charge is found in the outlying parts of atoms hereabouts,
and positive charge is found in the central parts. O-language has the
resources to say so,  and we may assume that the postulate mentions whatever
it takes \textit{to break} such symmetries. Thus the theoretical roles of positive
and negative charge are not purely nomological roles; \textit{they are locational
roles as well}.  [my italic]
\end{quote}

 But Healey argues that, in  gauge theories, even  this Lewisian strategy is bound to be plagued by under-determination.  I interpret Healey as saying that  one can   functionally specify a representational convention for a  spacetime metric using only O-terms, but cannot specify a representational convention for a gauge potential. 
More precisely, here is Healey's argument that the functionalist methodology applies so as to single out spacetime metrics, but not to single out gauge potentials:
\begin{quote}
The idea seems to be to secure unique realization of the terms [e.g. the gauge potential $A$...] 
in face of the assumed symmetry of the fundamental theory in which they figure by adding one or more sentences [namely, $S$] stating what might be thought of as ``initial conditions'' to the laws of that theory 
[...] to break the symmetry of how these terms figure in  $T$. They would do this \textit{by applying further constraints} [...namely, $S$...] 
Those constraints would then fix the actual denotation of the  
[...symmetry-related terms...] in $T$ so that, subject to these further constraints, $T$ is uniquely realized. [...But]  \textit{The gauge symmetry of the theory would prevent us from being able to say or otherwise specify which among an infinity of distinct distributions so represented or described is realized in that situation. This is of course, not the case for general relativity.}  \citep[p. 93]{Healey_book} [my italics]
\end{quote}

But I would ask: Why is this ``of course not the case for general relativity''? And why does Healey see here a contrast between general relativity and gauge theory?    
 I do not see in this entire passage an attempt to draw a distinction regarding Leibniz-equivalence and Sophistication about symmetries between the two cases.
I think the interesting question being alluded to here is whether we can use features of the world around us to single out  a unique model of the theory by its  unique features. 

This other question is more interesting because, in practice, we \emph{do} select some model over others when we represent a given physical situation, and therefore we must somehow `break the symmetry' between all of the models. 
In other words, if we accept that the basic ontology is symmetry-invariant, we are faced with a mystery:  how do we  select particular representatives of the ontology---i.e. models---over others? Under what conditions could we be justified in choosing for the gauge potential one spacetime distribution over another?
 Selecting such a representative  involves  a tension between: (a) a structural construal of physical properties---as  ones that are invariant under the symmetries in question---and (b) in practice  selecting unique   representative distribution, among the infinitely many isomorphic representatives of the same situation. At first sight, these two requirements, (a) and (b), are inimical, if not contradictory, for (a) implies we can have no physical guidance for accomplishing (b)! 
 
 Getting representational conventions off the ground of course depends on a positive resolution of this paradox. 
 For though no particular choice is mandatory, each must be based on something: physical features, indexicals, ostension, etc.

Thus I take the more interesting interpretation of Healey's passage here to be that this `singling out' of particular models is possible for gravity, but not for gauge theory, where the choice is unthethered from any feature of the physical world: it is entirely arbitrary, according to him. If this were so---if there was  nothing contingent on which we could base our representational conventions for gauge theory---we would indeed be more motivated to seek out reduction or elimination for  gauge theory than for gravity.\footnote{And indeed, later in the book Healey uses this distinction as a motivation for seeking  a different, symmetry-invariant ontology of gauge theory, based on holonomies.} Answering this question is thus crucial in order to clarify the scope of representational conventions.

\subsection{Refuting Healey's alleged distinction using representational conventions}\label{sec:antiheal}
\emph{Contra} Healey,  I say that having some physical ``hook'' with which to choose representatives does \emph{not} imply that we are breaking the symmetry at a  fundamental level. Different choices of representational conventions are equally capable of representing a given state of affairs; some may just be  more cumbersome than others.  Or perhaps some representations obscure matters for the purposes
at hand, even while they may shine a light on  complementary aspects of that state of affairs. 
 And, in this sense,  by choosing different conventions we  shift our focus to different features of the world, according to our interest. 
 
 So  we  construct a particular representative of the gauge potential as fulfilling a given role.  We resolve the tension between (a) and (b) described above, with explicit examples; by,  in Healey's words: `\textit{breaking} the symmetries', by providing `\textit{further constraints}'.
 
  Note to begin with that, according to Healey's standards, we are justified in including in our O-vocabulary all the `locational roles', which  describe contingent, happenstantial facts about  `where and when' specified events happen; and which I will loosely interpret as `referring to spacetime'. Thus I free myself to include in Lewis\rq{} O-vocabulary, and thereby use in the specification of the roles, the differential geometry of spacetime. In short, I will assume reference to spacetime is `old' or already understood.

In the simple example of electromagnetism, we require the model  to satisfy certain relations among the parts of the field. For example, in our hierarchy of extra-empirical theoretical virtues, we could place Lorentz covariance very highly (cf. \cite{Mulder_AB} and  \cite{Mattingly_gauge} for advocacy of this criterion and choice of gauge) and therefore prefer an explicitly Lorentz-covariant choice of convention.

 The only extra constraints that we have imposed in this equation, namely, that the spacetime divergence of the particular representative of $\omega$ vanishes, use only  the O-vocabulary that Healey would grant us.\footnote{There is a subtle point here: we are in effect assuming the active-passive correspondence of \cite[Sec. 5]{Samediff_0}, to say that fixing a choice of section of the bundle is equivalent to picking out a unique model within an isomorphism class. \label{ftnt:active_passive}} 

Of course we still have the freedom to change the section $\sigma$, but a different section would not satisfy the original  condition that    uniquely specified $\sigma$. 
 The representational convention cannot be empirically \emph{mandated}, since, by assumption, the symmetries leave all empirical matters invariant, in both the gravitational and the gauge cases. 
 
  For example, each spacetime point may have \emph{many} distinct, uniquely individuating properties. But this is nothing special to  theories of physics that admit symmetries: it is just like the tallest woman in Sweden, who can also be individuated by her fingerprints, as well as by her height and location. Likewise, `relationism' is a broad notion, admitting many physical bases for describing the world.

 No, each choice is at most suggested by being \emph{suitable} for certain types of questions one might want to ask about the system.
Different choices merely represent different lenses through which we capture the invariant structure of the states: the structure-tokens. Because these choices are not mathematically underpinned, it is apt to call them `representational\rq{}, and not `instantiation\rq{} conventions (cf. footnote \ref{ftnt:instantiation}). 
 
We should not be disappointed by this lack of uniqueness. It should be seen instead as a \emph{flexibility} that is also explanatory. Just as it is easy to explain the Larmor effect by a Lorentz boost between different frames,  we use different gauges to explain that a given process in quantum electrodynamics involves just two physical polarisation states or that it is Lorentz invariant. In both the special relativistic and the gauge scenarios, two different `frames'---for quantum field theory, the temporal and Lorenz gauge---are necessary to more easily explain two different aspects of a given phenomenon.  In the  words of \citet[p. 1]{Tong_gt}:
\begin{quote}
 The [gauge] redundancy allows
us to make manifest the properties of quantum field theories, such as unitarity, locality,
and Lorentz invariance, that we feel are vital for any fundamental theory of physics
but which teeter on the verge of incompatibility. If we try to remove the redundancy
by fixing some specific gauge, some of these properties will be brought into focus,
while others will retreat into murk. By retaining the redundancy, we can flit between
descriptions as is our want, keeping whichever property we most cherish in clear sight.
\end{quote}

  In sum,  the representational convention can be seen as specifying a functional role, using different  pragmatic, explanatory, and theoretical requirements. The choice of gauge that it implies finds \lq{}sufficient reason\rq{} in the satisfaction of this role; there is no unwarranted breaking of gauge-symmetry. 
   This construction thus explicitly contradicts the letter of Healey's under-determination argument. 
   This concludes  my response to what I called Healey's `interesting challenge': showing that intra-theoretic resources enable us to choose representational conventions and  pick out representatives. 

\section{Representational conventions resolving Worry (2b)}\label{sec:rep_app}\label{sec:counterfactual}

 I take Worry (2a)---the issue described in Section \ref{subsec:skeptic_why2}: that Sophistication left open the matter of providing sufficient and tractable conditions of identity for \emph{structure-tokens}---to have been now completely resolved in Section \ref{sec:rep_conv}, exemplified in Section \ref{sec:examples}, and clarified in Section \ref{sec:Healey}.

But we still have not explicitly addressed how representational conventions also resolve  Worry (2b). This is the aim of this Section.

    To recap, the second unresolved issue is that (cf. Section \ref{sec:challenges}):
\begin{enumerate}[2b.]
\item The strictly qualitative description of  theories eliminates any primitive identity across physical possibilities, or, equivalently, across different isomorphism-classes. But such identities were useful in order  to  express counterfactuals. So how are counterfactuals expressed   qualitatively?
\end{enumerate}
 For concreteness, and since most of the literature focuses on Worry (2b) in the case of spacetime, I will here also focus in that application; but most conclusions will also apply to the gauge case (by replacing `points\rq{} by `internal frames\rq{} of a vector space).

More specifically,  I will address  Worry(2b)  through another Lewisian topic:  countepart theory. Counterpart theory is the philosophical doctrine, due to David Lewis (\cite{Lewis1968}, \cite[p. 38-43]{Lewis1973}  and  \cite[Ch. 4]{Lewis1986}) that any two objects---in particular, spacetime points---in two different possibilities (in philosophical jargon: possible worlds) are never strictly identical. They are distinct, though of course  similar to each other in various, perhaps many, respects.  Representational conventions, specified by extra-empirical criteria as in Section \ref{sec:Healey}, can be construed as picking out which respects are important for the issue at hand.\footnote{This idea of using gauge-fixings as counterpart relations was introduced in \citep{Gomes_PhD} and developed in fullness in \cite{GomesButterfield_hole2}; here I will give only a sketch.}  

As to counterfactuals,  what makes true a proposition that the object $a$ could have had property $F$ (though in fact it lacks $F$) is not that in another possible world, $a$ {\em itself} is $F$, but that in another possible world, an object appropriately similar in certain respects to $a$, is $F$. That object is called the {\em counterpart}, at this other possible world, of $a$. 

Summing up the Worry and the strategy to answer it: though denying trans-world identity fits sophistication's denial of any sort of primitive identity, counterparthood furnishes a relation that Sophistication lacks, allowing us to discuss counterfactuals.  Each representational convention via gauge-fixing provides explicit and invariant local counterpart relations between  non-isomorphic models. It does this most clearly  in its guise as a relational reference frame.
That is because, as I said in Section \ref{sec:ref_frames}, relational reference frames  make precise a frequent claim found in the literature on Sophistication about spacetime: that `spacetime points can only be specified by their web of relations to other points\rq{}. Each gauge fixing weaves a web of relations that is rigid with respect to locations but maximally loose with respect to the physical content of those locations.  Thus we can specify spacetime points in different possibilities---in non-isomorphic models---by their location within the   web of relations provided by the corresponding relational reference frame, and compare the values that other physical quantities take therein.\footnote{In this Section, since we will be comparing objects across different possibilities, and using different comparisons, I prefer to talk about  \emph{specifying} a point or region, rathern than individuating it, since the latter term\rq{}s  connotation of identity is stronger than is required here. See footnote \ref{ftnt:specify}.}

 In Section \ref{sec:formalism_counter} I provide the basic formalism for conceiving of counterpart relations as given by representational conventions and discuss  how changing representational conventions affects the counterpart relations; in Section \ref{sec:snag} I describe obstructions to construing counterpart relations in this manner---obstructions due to homogeneous models.

\subsection{Basic formalism}\label{sec:formalism_counter}
 Mathematically, a counterpart relation can be seen as a group element $g\in \G$ that relates two different models $\varphi_1$ and $\varphi_2$, not necessarily isomorphic. For instance, given two models for vacuum general relativity, $\langle M, g^1_{ab}\rangle$ and $\langle M, g^2_{ab}\rangle$,  a diffeomorphism $f\in \Diff(M)$ will give us a counterpart relation between the spacetime points of $M$ in each of the two models. Similarly, a vertical automorphism between two models of Yang-Mills theory: $\langle P, \omega_1\rangle$ and $\langle P, \omega_2\rangle$, will give us a counterpart relation for $P$. 
 
 Given a section $\mathcal{F}_\sigma$, as described in Section \ref{subsec:gf}, we have just such an element: the dressing $g_\sigma: \F\rightarrow \G$, given in Equation \eqref{eq:gauge-fixing}. Using this convenient mathematical operator, the counterpart relation between $\varphi_1$ and $\varphi_2$ is given by:
 \be\label{eq:counter} \text{Counter}_\sigma(\varphi_1, \varphi_2):=g_\sigma(\varphi)g_\sigma(\varphi_2)^{-1}.
 \ee
Generally, i.e. for models that are not along the section,  the relation is given by  the group element that takes the first model down to the gauge-fixing section and then back up towards the second model.  In particular, we see that the counterpart relation between any two states that already lie at the gauge-fixing section is just the identity: they are both already in their preferred representation relative to the reference frame that is associated to the gauge-fixing. And by being associated to a gauge-fixing, we don\rq{}t limit the physical content that we seek to compare via  counterparthood. 
Thus each section gives a qualitative, i.e. diffeomorphism-invariant, counterpart relation between the dressed spacetime points of distinct physical possibilities.   
 
 Due to  covariance  (see \eqref{eq:cov_g}),
 \begin{equation}\label{covarycpart} 
{\rm Counter}_{\sigma}(\varphi^g_1, \varphi^g_2) = g^{-1}g_\sigma(\varphi_1)(g^{-1}g_\sigma(\varphi_2)^{-1}= g^{-1}\,{\rm Counter}_{\sigma}(\varphi_1, \varphi_2)g\; .
\end{equation}
 Moreover, two models that lie in the same orbit will always be related by the unique isomorphism that connects them. That is: if $\varphi_2 =  \varphi^g_1$, then ${\rm Counter}_{\sigma}(\varphi_1, \varphi_2) = g$, even if neither $\varphi_1$ nor $\varphi_2$ lie in the section $\sigma$. 
For with the definitions above,
\begin{equation}\label{same}
{\rm Counter}_{\sigma}(\varphi_1, \varphi^g_1) = g_{\sigma}(\varphi_1)(g^{-1}g_{\sigma}(\varphi_1))^{-1}=g
\end{equation} 
Thus, for example, in the spacetime case, if the unique counterpart of $p$ in model $\langle M, g_{ab}\rangle$ is $q$ in model $\langle M, g'_{ab}\rangle$, then the counterpart of $p$ in model $\langle M, f^*g'_{ab}\rangle$ will be $f(q)$. And this particular property is independent of which convention $\sigma$ we choose.

In other words,   though there are several distinct choices of counterparts, each choice must identify the same spacetime points across isomorphic models.  In the spacetime context, this  is essentially \citet[p. 77]{Field_soph}'s observation that 
\begin{quote}
“individuation of objects across possible worlds” is sufficiently tied to their
qualitative characteristics so that if there is a unique 1-1 correspondence
between the space-time of world A and the space-time of world B that
preserves all geometric properties and relations (including geometric relations
among the regions, and occupancy properties like being occupied by
a round red object), then it makes no sense to suppose that identification of
space-time regions across these worlds goes via anything other than this
isomorphism. \end{quote}
Indeed, this criterion, that we can here call `the drag-along' interpretation of isomorphic models (cf. \cite{GomesButterfield_hole}), was the centerpiece of \cite{Weatherall_hole}: the paper that sparked a recent surge of interest in the hole argument in general relativity.\footnote{ \cite{Weatherall_hole, BradleyWeatherall_hole}  associate the `the drag-along' idea less with Field and more with \cite{Mundy1992}, who distinguishes a theory's synthetic language---that is only able to expresses qualitative, non-singular facts---from a theory's metalanguage, in which we are able to talk about points singularly, i.e. without a definite description (see \cite[Sec. 3.2a]{Samediff_1a}).  But the gist is the same: isomorphisms--- expressed in the metalanguage in Mundy's case---will map objects singled out by the same description into each other. Here is  \citet[p. 520]{Mundy1992}:
\begin{quote}
Thus an element $p$ of $M$ satisfies a description $D$ of $L_R$ [i.e. a qualitative description] in $S$ iff the corresponding element $f(p)$ satisfies $D$ in $Sf$: $p$ and $fp$ occupy the same structural roles in these two isomorphic models, so every statement true of $p$ in $S$ is true of $f(p)$ in $Sf$. Therefore, since a theory identifies elements of its domain only by descriptions expressed in its language, two such models describe the same theoretical world. 
\end{quote}}
 And here we see that this criterion is satisfied by our notion of qualitative counterpart relations using representational conventions. But we do not always have universal, unique, counterpart relations, and so Field's criterion  does not always apply. We now turn to this. 

\subsection{A snag for counterparts but not for conventions}\label{sec:snag}
Given any representational convention via gauge-fixing, in accord with  Field's criterion, above, as long as  the web of properties and relations in each of our models is sufficiently complex, it suffices to uniquely specify points across models.

But if this web is not sufficiently complex, the specification will fail. 
 To be blunt: we cannot stipulate a convention for some region  in spacetime or frame in a vector space if there aren't enough  specific features to single them out.

 As described in Appendix \ref{app:PFB} (see also end of Section  \ref{sec:rep_gen}),  models that are `too homogeneous\rq{}, or symmetric, have stabilizers, which represent a degeneracy in the web of relations that they can furnish. 
That is, on homogeneous states, $\tilde \varphi$, a representational convention---seen as a map $\sigma:[\Phi]\rightarrow \Phi$---will fail to uniquely specify the $g_\sigma$ of \eqref{eq:gauge-fixing}; $g_\sigma$ fails to satisfy the property of \emph{Uniqueness} of Section \ref{subsec:gf}. There are particular $\tilde g\in \G$ such that $\tilde \varphi^{\tilde g}=\tilde\varphi$, and so, in particular, we cannot fulfill  the covariance Equation \eqref{eq:cov_g}.  



 More broadly, this type of obstruction to individuation due to  homogeneity  is well known, and related to \cite{Black_PII}'s criticism of Leibniz's Principle of the Identity of Indiscernibles---the principle that motivates Leibniz equivalence (see \citep[Sec. 3.3]{Pooley_draft}  for a thorough exposition). The problem now is that, if, with \cite{Weatherall_hole}, one is tempted to interpret Field\rq{}s criterion very strongly,  identifying elements of one model with those of the other by the isomorphism relation, then the transitivity of identity would lead to  the disastrous conclusion that homogeneous spacetimes---whose isometry group can take any spacetime point to any other---`truly' contain only one point (cf. \cite{Wuthrich_abysmal})!

 Nonetheless, precisely because the degeneracy in $g_\sigma$ acts trivially on the respective models, it does not yield any degeneracy about which models belong to the gauge-fixing surface. And this is consistent with Sophistication: 
 while homogeneous models  fail  to satisfy Field's criterion, by failing to satisfy its antecedent,  Leibniz equivalence is holistic and  survives unscathed. For  it merely states that physical possibility is entirely determined by qualitative properties and relations \emph{of each entire world or model}, and even a homogeneous world can be qualitatively described perfectly well  (cf. also the related \citet{Muller_abysmal}'s counterargument to \cite{Wuthrich_abysmal}).\footnote{For instance, maximally homogeneous spacetimes are singled out by having the maximal number of Killing vector fields: a completely qualitative description.} 

In sum, for very homogeneous states, $g_\sigma$ is not well-defined,  so the counterpart relation \eqref{eq:counter} is no longer valid and will not recover Field's criterion. 
Nonetheless, the interpretation of isomorphic models through Sophistication and Leibniz equivalence survives unscathed and is carried through to  representational conventions, which are still well-defined.

\section{Summing up }\label{sec:conclusions}
\label{sec:pow_rep}
In this Section, I conclude, by first, in Section \ref{sec:resolution},  offering a very brief summary.  Then, in Section \ref{sec:obj}, I list a couple of idealizations involved in representational conventions, which I have so far skatted over. 

 \subsection{Representational conventions conceptually summarised}\label{sec:resolution}

In both the gravitational and the gauge cases, there are many  ways  in which we can  instantiate qualitatively identical patterns of e.g. parallel transport; each of these  ways correspond to  one of many \emph{isomorphic models} of the theory. But we can pick a particular representation by resorting to a \emph{representational convention}.

  Representational conventions in the context of these theories will introduce a  notion of `relationism' into the formalism. This is not necessarily the notion that is usually attributed to Mach, Poincar\'e, Barbour, etc. All I mean by it is that it requires relative values of fields. 
So a representational convention can be seen as  way to completely convey  the particular physical state by stating what are the values of meaningful, but {\em relational}, quantities for it. Each choice of representational convention can be identified as a choice of a set of relations used to describe physical content without redundancy, i.e. satisfying Definition \ref{def:rep_conv}. 

One main difference between mine and the traditional notion of relationism is that I explicitly admit multiple choices of sets of relations:  one set can only be  \emph{pragmatically} preferred to others.  Thus, when Einstein famously resolved his questions about the `hole argument\rq{} by appealing to `point coincidences\rq{} (cf. \cite{Giovanelli2021}), he assumed there was no further question as to what these points were coincidences of. But, as we have seen, different choices of matter fields or composites can yield very different physical descriptions of  given state of affairs: very different coincidences, which equally well coordinitize a given region of spacetime. 
 
  For instance,  in non-relativistic particle mechanics, the relational (i.e. dressed) variables could be the set of all constituent particles' distance to a center of mass and some preferred orientation therein (e.g. as a function of an anisotropic distribution of mass). In field theory, the set of relations  could be a  non-local composite of all the field, because they often  involve comparisons of values of  fields at different spacetime points. One such composite dresses a `bare\rq{} electron with its Coulombic `tail\rq{} \citep{Dirac:1955uv}.
  
  Using a fixed choice of representational convention we have a tractable criterion for when two generic models instantiate the same  gauge-invariant, or qualitative (in the jargon of metaphysics), pattern of parallel transport. Namely, when the values of the dressed fields  numerically match.  We thus reduce matters of structural identity to numerical coincidence. This answers Worry (2a).

As to Worry (2b), I only resolved it with representational conventions via gauge-fixing using the more specific  Definition \ref{def:rep_conv_n}. 
 For we can  construct from such a gauge-fixing relational observables called \lq{}dressed variables\rq{}: invariant quantities that satisfy the same conditions than the gauge-fixing section.

These variables can be understood relationally: as values of certain quantities expressed relative to some physical subsystem: these we called \lq{}relational reference frames\rq{}. Using them, we are able to characterise  e.g. regions of space  via their invariant attributes, but without severely limiting the possible physical content for those regions. Thus we give a new mathematical gloss on  counterpart theory. Each representational convention  gives us a particular inter-world counterpart relation between `qualitatively different patterns\rq{}, i.e. between non-isomorphic models. Roughly, in this language,  the idea is that the qualitative relations singled out by one convention provide similarity relations across physical possibilities. So questions such as: `what is the counterpart of event $p$ of our world  in a qualitatively different pattern?' make sense under a representational convention, since there will be a unique point $q$ at a different world that shares the subset of  properties that are ``selected as salient'' by the convention.

  And finally, there is the matter of which convention to choose. To pick a convention we must employ intra-theoretic resources by the lights of which the convention  is appropriate. Isomorphisms of a theory can then be understood as  changes in the (context-dependent) relations and properties we choose to  instantiate the structure-tokens.

  In sum,  representational conventions can be construed as (non-unique) choices of a relational set of quantities that:
\begin{enumerate}[(I)]
\item provide  tractable,  complete characterizations of  structure-tokens, i.e. of the  structure  that, given a fixed notion of isomorphism, is:  common to each class of isomorphic models, and different for different classes of isomorphic models. They give a tractable criterion of individuation of structure-tokens. 
\item  allow us to describe qualitative counterpart relations for points, regions, frames, etc, across isomorphic and non-isomorphic models, while  entirely denying any primitive---i.e. convention-independent---identification   across models. 
\end{enumerate}
And so, to complement my first two Desiderata for Sophistication of \cite{Samediff_1a}, I here introduce Desideratum (iii): that the theory's symmetries admit representational conventions. Fulfillment of this Desideratum resolves the first two Worries of Section \ref{sec:challenges}, left over from \cite{Samediff_1a}. But is Desideratum (iii)  a non-trivial requirement?  

\subsection{Further objections.}\label{sec:obj}
Representational conventions are limited in a way  which we have only  touched on in passing so far. Take De Donder variables as defined by four massless scalars satisfying a wave equation (cf. Equation \eqref{eq:Komar_coords}). Even if their domain is a generic spacetime, that spacetime will (generically) not be homeomorphic to $\RR^4$. As we know, there are topological limitations to the existence of such maps, and there might be other obstructions as well. For instance, a spacetime may simply not have any point at which all of the Komar invariants vanish. The upshot is that representational conventions aren\rq{}t usually global, either in spacetime or in $\F$. 

Recall from Section \ref{subsec:skeptic_why2} that individuating structure-tokens is a very common goal for theorists working in quantum gravity, and in the (non-perturbative) quantization of Yang-Mills theory. Indeed,  the global limitations of representational conventions  are  major obstacles to achieving this goal. For instance, in quantum gravity, the infamous ``problem of time'' is often portrayed as a problem of finding suitably general `internal clocks' (cf. e.g. \cite[Sec. 3.4]{Isham_POT}).
\footnote{Similarly, in general more than one representational convention is required to cover the entire spacetime manifold. We reiterate that the introduction of transition functions is generally necessary because global sections do not exist, also in the gauge case,  unless the bundle is trivial, i.e. unless $P= M \times G$ {\it globally} not just locally. In the trivial case, and only in the trivial case, all transition functions can be trivialized to be the identity, i.e. $\t_{\beta\alpha} = g_\beta g_\alpha^{-1}$ for some choices of $g_\alpha$'s. 
  This fact has important consequences for the treatment of subsystems for theories with local symmetries, since it obliges us to consider field-dependent transition functions. This is the reason   representational conventions were originally introduced for subsystems in gauge theory and general relativity.}

This is why Desideratum (iii)--- admitting a representational convention---is highly non-trivial. In fact, it is not perfectly satisfied even for   non-Abelian Yang-Mills and general relativity, due to the `Gribov obstruction\rq{} (see Appendix \ref{app:PFB}).  But it is satisfied in the Abelian case, and its general features should be present even with the caveats above.

 In the original treatment of \cite{GomesRiello2016, GomesRiello2018, GomesHopfRiello, GomesRiello_new}, one avoids the problem of Gribov ambiguities by introducing a more flexible tool than representational conventions, called a `relational connection-form\rq{}. In brief, this is a distribution of infinitesimal gauge-fixing surfaces, with appropriate covariance properties. These connection forms generalize parallel transport in principal fiber bundles to parallel transport in the space of models, where the gauge group forming the orbits is the entire group of gauge transformations. In that context, connection-forms provide counterpart relations between members of any one-parameter sets of models---i.e. along `histories\rq{} of models---and so generalise Barbour\rq{}s `best-matching\rq{} (see e.g. \cite[Ch. 4]{Flavio_tutorial}).  Heuristically, this can be understood as picking out, along neighboring orbits, the models that are closest, according to some choice of gauge-invariant notion of `closeness\rq{} that applies to entire models.
 
 The transversal, covariant distribution defined by a relational connection form is not necessarily integrable into `slices' (see Appendix \ref{app:PFB}), which means it usually has curvature in $\F$. In this case, since history-dependent, counterpart relations given by connection-forms will not generally satisfy `Field\rq{}s criterion\rq{}  (see \citep{GomesHopfRiello} for a pedagogical introduction).\footnote{This more flexible tool  (which was technically developed in \cite{GomesRiello2016, GomesRiello2018, GomesHopfRiello, GomesRiello_new} has been conceptually appraised in \cite{GomesStudies, Gomes_new} and \cite[Appendix A]{GomesButterfield_hole2}. }

\subsection*{Acknowledgements} I would like to thank Jeremy Butterfield for many detailed comments on several versions of this paper. I would like to thank  the British Academy for financial support. 

 \begin{appendix}
 \section*{APPENDIX}
 
 \section{Infinite-dimensional principal fiber bundles}\label{app:PFB}\label{ftnt:slice}
 
 In the case of field theories, such as general relativity and Yang-Mills, even admitting a free and proper action of an infinite-dimensional Lie group on an infinite-dimensional smooth manifold, it is not guaranteed that a local product structure exists everywhere: so $\F$ is not necessarily a \emph{bona-fide} infinite-dimensional principal $\mathcal{G}$-bundle, i.e. $\mathcal{G}\hookrightarrow\F\rightarrow \F/\mathcal{G}$.  The obstacle is that there are special states---called \emph{reducible}---that have stabilizers, i.e. elements $\tilde g\in \G$ such that  $\tilde\varphi^{\tilde g}=\tilde \varphi$. Moreover, the statement holds for the entire group orbit, as it is easy to show that for  some $\tilde\varphi':=\tilde\varphi^g\in\mathcal{O}_{\tilde \varphi}$, 
$$\tilde\varphi'^{g^{-1}\tilde g g}=\tilde\varphi^{\tilde g g}=\tilde\varphi^g=\tilde\varphi'$$
 and so all the elements of the orbit are also reducible (with stabilizers related by the co-adjoint action of the group). And so  orbits are of `different sizes', and not isomorphic to the structure group. Nonetheless, there is a generalization of a section, called a slice, that provides a close cousin of the required product structure (see \cite[Sec. 2.1, and footnotes 4,5]{GomesButterfield_hole2}). As has been shown using different techniques and at different  levels of mathematical rigour,  \cite{Ebin, Palais, Mitter:1979un, isenberg1982slice, kondracki1983, YangMillsSlice, Slice_diez} both the Yang-Mills configuration space and the configuration space of Riemannian metrics (called $\mathrm{Riem}(M)$), admit slices. These slices endow the quotient space with a \emph{stratified} structure. That is, the  space of models can be organised  into orbits of models that possess different numbers of  stabilizers; with the orbits with more stabilizers being at the boundary of the orbits of models with fewer stabilizers. For each stratum, we can find a section and form a product structure as in the standard picture of the principal bundle.

For both general relativity and non-Abelian gauge theories,   reducible configurations   form a meagre set.   \textit{Meagre}  sets are those that arise as countable unions of nowhere dense sets. In particular, a small perturbation will get you out of the set (and this is true of the reducible states in the model spaces of those theories, according to the standard field-space metric topology  (the Inverse-Limit-Hilbert topology cf. e.g. \cite{kondracki1983, fischermarsden}). In this respect, Abelian theories, such as electromagnetism, are an exception: {\it all} their configurations are reducible,  possessing the constant gauge transformation as a stabilizer. 

And apart from this obstruction to the product structure---i.e. even if we were to restrict attention to the generic configurations in the case of non-Abelian field theories---one can have at most a \emph{local} product structure: no representational convention, or section, giving something like \eqref{eq:doublet}, is global (this is known as the  \emph{Gribov obstruction}; see \cite{Gribov:1977wm, Singer:1978dk}). Unfortunately, the space of Lorentzian metrics is not known to have such a structure: it has only been shown for the space of Einstein metrics that admit a constant-mean-curvature (CMC) foliation.

In the infinite-dimensional case, both the dimension and the co-dimension of a regular value surface can be infinite, and it becomes trickier to construct  a section: roughly, one starts by endowing $\F$ with some  $\G$-invariant (super)metric, and then finds the orthogonal complement to the orbits, $\cal{O}_\varphi$, with respect to this supermetric. But here the intersection of the orbit with its orthogonal complement cannot be assumed to vanish, as it does in the finite-dimensional case. Nonetheless, in the cases at hand, that intersection is given by the kernel of an elliptic operator, and one therefore can invoke the `Fredholm Alternative' (see \cite[Sec. 5.3 and 5.9]{Trudinger}) to show that that intersection is at most finite-dimensional, but generically is zero, and thus the generic orbit has the `splitting' property: the total tangent space decomposes into a direct sum of the tangent space to the orbit and its orthogonal complement. Now we must extend the directions transverse to the orbit, so as to construct a small patch that intersects the neighboring orbits only once. But the space of Riemannian  metrics is a cone inside a vector  space, so it is not even affine and we cannot just linearly extend the directions normal to the orbit and hope for the best. And so we extend   the normal directions by  using the Riemann normal exponential map with respect to the supermetric (cf. \cite{Gil-Medrano}), and thus conclude that, for a sufficiently small radius, the resulting submanifold is transverse to the neighboring orbits and   has no caustics. Finally, to show that this `section' is not only transverse to the orbits, but indeed that it intersects neighboring orbits only once, the orbits must be embedded manifolds, and not just local immersions: this is guaranteed if the group action is proper, cf. \cite{Ebin}. 

\section{The holonomy representation}\label{app:hol}
We can assign a complex number  (matrix element in the non-Abelian case) $hol(C)$ to the oriented embedding of the unit interval: $C:[0,1]\mapsto M$, by integration of a phase:
\be\label{eq:hol} hol_C(A) =\exp{i\int_C A}.\ee
Under a gauge transformation, we obtain: 
\be hol_C(A^g)=g^{-1}(C(0))hol_C(A) g(C(1)). 
\ee

If the endpoint of $C_1$ coincides with the starting point of $C_2$, we define the composition $C_1\circ C_2$ as, again, a map from $[0,1]$ into $M$, which takes $[0,1/2]$ to traverse $C_1$ and $[1/2, 1]$ to traverse $C_2$.  The inverse $C^{-1}$ traces out the same curve with the opposite orientation, and therefore $C\circ C^{-1}=C(0)$.
Following this composition law, it is easy to see from \eqref{eq:hol} that 
\be\label{eq:loop_com} hol(C_1\circ C_2)=hol(C_1)hol(C_2),\ee with the right hand side understood as complex multiplication in the Abelian case, and as composition of linear transformations, or  multiplication of matrices, in the non-Abelian case.
For both Abelian and non-Abelian groups, given the above notion of composition, holonomies are conceived of as smooth homomorphisms from the space of loops into a suitable Lie group. One obtains a representation of these abstractly defined holonomies as
holonomies of a connection on a principal fiber bundle with that Lie group as structure group; the collection of such holonomies carries the same amount of information as the gauge-field $A$ (cf. \cite[Sec. 3]{Belot1998} for a philosophical exposition). However, only for an Abelian theory can we cash this relation out in terms of gauge-invariant functionals. That is, while \eqref{eq:hol} is gauge-invariant, the non-Abelian counterpart (with a path-ordered exponential), is not. For non-Abelian theories the gauge-invariant counterparts of \eqref{eq:hol} are Wilson loops, see e.g. \citep{Barrett_hol}, 
$ W(\gamma):=\text{Tr}\, \mathcal{P}\exp{(i\int_\gamma A)}
$,
where one must take the trace of the (path-ordered) exponential of the gauge-potential. It is true that all the gauge-invariant content of the theory can be reconstructed from Wilson loops;  (see also \cite{RosenstockWeatherall2016c}, for a category-theory based derivation of this equivalence). But,  importantly for our purposes, it is no longer true that there is a homomorphism from the composition of loops to the composition of Wilson loops. That is, it is no longer true that the counterpart \eqref{eq:loop_com} holds,  $W(\gamma_1\circ\gamma_2)\neq W(\gamma_1)W(\gamma_2)$.  The general composition constraints---named after Mandelstam---come from generalizations of the Jacobi identity for Lie algebras, and depend on $N$ for SU($N$)-theories; e.g. for $N=2$, they apply to three paths and are: 
\begin{eqnarray}\label{eq:Mandel}W(\gamma_1)W(\gamma_2)W(\gamma_3)-\frac12(W(\gamma_1\gamma_2)W(\gamma_3)+W(\gamma_2\gamma_3)W(\gamma_1)+W(\gamma_1\gamma_3)W(\gamma_2)\nonumber\\
+\frac14(W(\gamma_1\gamma_2\gamma_3) + W(\gamma_1\gamma_3\gamma_2) = 0.\end{eqnarray}
\end{appendix}

\bibliographystyle{apacite} 
\bibliography{references3}
\end{document}